\documentclass[prd,aps,twocolumn,superscriptaddress,reprint,nofootinbib,preprintnumbers]{revtex4-2}
\usepackage{graphicx}
\usepackage{hyperref}
\usepackage{color}
\usepackage{amsmath, amsthm, amssymb, amsfonts, latexsym}
\usepackage{bm}
\usepackage{bbold}
\usepackage[dvipsnames]{xcolor}
\usepackage{physics}

\newcommand{\nopi}{\ensuremath{\pi\hskip-0.40em /}}
\newcommand{\eftnopi}{EFT$_{\nopi}$}

\def\beq{\begin{equation}}
\def\eeq{\end{equation}}

\usepackage{array}
\newcolumntype{P}[1]{>{\centering\arraybackslash}p{#1}}
\newcolumntype{M}[1]{>{\centering\arraybackslash}m{#1}}

\definecolor{darkblue}{cmyk}{1,0.4,0,0.3}
\definecolor{violet}{cmyk}{0,1,0,0.2}
\hypersetup{colorlinks, bookmarksnumbered, citecolor=darkblue, linkcolor=darkblue, pdfstartview=FitH, urlcolor=darkblue, linktocpage, pdftitle={Quantum Resources and Wigner Symmetry in Nucleon-Nucleon Scattering from Effective Field Theory}, pdfauthor={Ian Low, Thomas R. Richardson, Sokratis Trifinopoulos}}

\newcommand{\oneS}{{{}^{1}\!S_0}}
\newcommand{\threeS}{{{}^{3}\!S_1}}
\newcommand{\onePone}{{{}^{1}\!P_1}}
\newcommand{\Ptwo}{{{}^{3}\!P_2}}
\newcommand{\Pone}{{{}^{3}\!P_1}}
\newcommand{\Pzero}{{{}^{3}\!P_0}}
\newcommand{\PJ}{{{}^{3}\!P_J}}
\newcommand{\oneP}{{{}^{1}\!P_1}}

\newcommand\couplingOneZero{C_{1\cdot 1}}
\newcommand\couplingThreeOne{C_{G\cdot G}}
\newcommand\couplingSixOne{C_{G \cdot G}'}
\newcommand\couplingOneOne{C_{\tau \cdot \tau}}

\newcommand\couplingThreeZero{C_{\sigma\cdot \sigma}}

\newcommand\couplingFiveZero{{\LRC}_{1\cdot \sigma}}

\newcommand\couplingSixZero{C_{\sigma\cdot\sigma}'}

\newcommand{\LRd}{\overset{\leftrightarrow}{\nabla}{}}
\newcommand{\LRC}{\overset{\leftrightarrow}{C}{}}

\newcommand{\calO}{\ensuremath{\mathcal{O}}}

\newcommand{\calA}{\ensuremath{\mathcal{A}}}

\newcommand{\calL}{\ensuremath{\mathcal{L}}}
\newcommand{\calM}{\ensuremath{\mathcal{M}}}

\newcommand{\calH}{\ensuremath{\mathcal{H}}}
\newcommand{\calE}{\ensuremath{\mathcal{E}}}
\newcommand{\calT}{\ensuremath{\mathcal{T}}}
\newcommand{\1}{\mathbb{1}}

\begin{document}

\preprint{CERN-TH-2026-138}
\preprint{INT-PUB-26-026}
\title{Quantum Resources and Wigner Symmetry in Nucleon--Nucleon Scattering from Effective Field Theory}

\author{ Ian Low}
\email{ilow@northwestern.edu}
\affiliation{Department of Physics and Astronomy, Northwestern University, Evanston, IL 60208, USA}
\affiliation{High Energy Physics Division, Argonne National Laboratory, Argonne, IL 60439, USA}
\author{ Thomas R.~Richardson}
\email{thomas.richardson@berkeley.edu}
\affiliation{Department of Physics, University of California, Berkeley, CA 94720, USA}
\affiliation{Nuclear Science Division, Lawrence Berkeley National Laboratory, Berkeley, CA 94720, USA}
\author{ Sokratis Trifinopoulos}
\email{sokratis.trifinopoulos@cern.ch}
\affiliation{Theoretical Physics Department, CERN, Geneva, Switzerland}
\affiliation{Physik-Institut, Universit\"at Z\"urich, 8057 Z\"urich, Switzerland}

\begin{abstract}
We study quantum resources in the spin degrees of freedom, such as entanglement, stabilizer magic, and non-local magic, in low-energy nucleon--nucleon scattering through next-to-leading order in pionless effective field theory.
Treating each nucleon spin as a qubit, we calculate the corresponding resource-generating powers of the scattering operator at generic center-of-mass momentum and scattering angle $\Theta$.
The analysis retains $S$- and $P$-wave channels generated by two-derivative contact interactions.
When the microscopic physics exhibits Wigner's $SU(4)$ spin-flavor symmetry, the neutron--proton amplitude becomes proportional to the identity gate in spin space and therefore generates no new resources after scattering, extending an observation previously made for leading-order $S$-wave scattering.
The same-nucleon channel remains resource-generating because constraints from identical particles project out part of the Hilbert space.
These results show how enhanced symmetries, partial-wave structure, and resource generation are intertwined in low-energy two-body scattering.
\end{abstract}

\maketitle

\section{Introduction}
\label{sec:intro}

Low-energy nucleon--nucleon ($NN$) scattering exhibits a set of striking approximate symmetries whose dynamical origin has long been a puzzle in nuclear physics.
The most prominent among these is the Wigner $SU(4)$ spin-flavor symmetry~\cite{Wigner:1936dx}, whose emergence at low energies can be traced to the large-$N_c$ expansion of QCD~\cite{Kaplan:1996rk, Kaplan:1995yg} (see Ref.~\cite{Richardson:2022hyj} for a review) and to the proximity of the two-nucleon system to unitarity~\cite{Mehen:1999qs}.

A complementary perspective on these emergent symmetries comes from applying concepts from quantum information (QI) theory to hadronic scattering.
Ref.~\cite{Beane:2018oxh} computed the entanglement power of the $S$ matrix for $NN$ scattering using leading-order (LO) pionless effective field theory (\eftnopi)~\cite{Kaplan:1996xu, Kaplan:1998tg, Kaplan:1998we, vanKolck:1998bw, Chen:1999tn}, valid for relative center-of-mass momenta $|\vb p|\ll m_\pi$ in the $S$-wave sector.
That work connected entanglement suppression to Wigner $SU(4)$ symmetry~\cite{Mehen:1999qs} and Schr\"odinger conformal invariance~\cite{Mehen:1999nd}, and suggested a possible role for QI diagnostics in nuclear effective field theory (EFT) power counting.
This connection was sharpened in Refs.~\cite{Low:2021ufv,Liu:2022grf}, which demonstrated that the unitary $S$ matrix can be written as a linear combination of operators acting on spin space that can be identified with the identity and SWAP quantum gates. Moreover, those works noted that realizing either gate structure in nature implies the emergence of enhanced symmetries.

The correlation between entanglement suppression and enhanced symmetries has since been extended across a broad range of hadronic systems, including pion scattering~\cite{Beane:2021zvo} and low-energy scattering of heavy mesons and higher-spin baryons~\cite{Hu:2024hex,Hu:2025lua,Hu:2025jne,Sone:2026jmo}.
Related connections between scattering amplitudes, entanglement, and more general $S$-matrix structures appear in Refs.~\cite{Low:2024hvn,Low:2024mrk,McGinnis:2025brt}.
Entanglement constraints have also been investigated in relativistic amplitudes of fundamental particles~\cite{Cervera-Lierta:2017tdt,Carena:2023vjc,Chang:2024wrx,Kowalska:2024kbs,Carena:2025wyh,Liu:2025iwh,Thaler:2024anb,Liu:2025bgw,Li:2026kha,Cao:2026aye}.
Beyond two-to-two scattering, Ref.~\cite{Low:2026oyf} found enhanced symmetries when the entanglement power of quantum many-body time evolution is extremized.

The quantum resources considered in this work probe three related structures carried by the outgoing two-spin state.
Entanglement measures nonseparability between the two spin qubits.
Magic, or nonstabilizerness, measures departure from the stabilizer sector that can be simulated efficiently with Clifford operations~\cite{Bravyi:2004isx,Emerson:2013zse}, while non-local magic isolates the part that cannot be removed by local basis changes~\cite{Qian:2025oit,Gargalionis:2026onv,Busoni:2026lvp}.
The corresponding resource powers, defined below, are state averages that quantify the ability of the scattering operator to generate these resources from appropriate resource-free input states.
Recent analyses have applied magic diagnostics to high-energy processes~\cite{Chernyshev:2024pqy,White:2024nuc,Aoude:2025jzc,Liu:2025qfl,Chang:2026nlo,Busoni:2025dns,Gargalionis:2025iqs,Nunez:2025xds}.
In nuclear physics, Ref.~\cite{Robin:2024oqc} studied magic and complexity fluctuations in low-energy nuclear and hypernuclear forces, and Ref.~\cite{Robin:2025ymq} later considered non-local magic diagnostics for two-particle scattering processes including low-energy $NN$ scattering.

The spin-entanglement properties of the neutron--proton ($np$) system
were also analyzed in Ref.~\cite{Bai:2023tey}, which computed the entanglement power and related spin-state resources  
using phenomenological two-nucleon potentials at arbitrary momentum and scattering angle.
Beyond the $S$-wave and including couplings breaking Wigner's $SU(4)$ symmetry, Ref.~\cite{Miller:2023ujx} made the observation that the transition matrix sometimes could maximize the spin entanglement of the outgoing particles in parts of the phase space.\footnote{This observation is not in conflict with that in Ref.~\cite{Beane:2018oxh}, which studied the entire $S$-matrix, not just the transition matrix. 
This distinction is an important one, which we elaborate in Sec.~\ref{sec:2a}.}
More recently, Ref.~\cite{Cavallin:2025kjn} studied how Wigner $SU(4)$, incorporated in potentials derived from chiral EFT, shapes spin entanglement in both $np$ and neutron--neutron ($nn$) scattering. Refs.~\cite{Witala:2025wvi,Witala:2025kat} identified polarized kinematic configurations in $np$ and neutron--deuteron scattering where strongly entangled final states can be produced experimentally.
Entanglement in few-nucleon systems has also been studied within \eftnopi\ and chiral EFT~\cite{Kirchner:2023dvg}.

A feature common to these studies is that either they operate in the $S$-wave approximation, where the resource information reduces to the two scattering lengths in the $^1S_0$ and $^3S_1$ channels, or they rely on phenomenological phase shifts that do not directly expose which low-energy coefficients (LECs) govern resource generation.
This complicates an attempt to understand the possible connection between the extremization of entanglement and emergent symmetries in low-energy nuclear physics.
A natural step towards resolving these questions is to study the 
structure of the quantum resources
generated by higher-order operators and higher partial waves in an EFT framework.
Specifically, we work through next-to-leading order (NLO) in \eftnopi\ and retain the two $S$-wave channels, $^{1}S_{0}$ and $^{3}S_{1}$, parametrized by their scattering lengths and effective ranges, and include the four leading $P$-wave channels, $^{1}P_{1}$, $^{3}P_{0}$, $^{3}P_{1}$ and $^{3}P_{2}$, through their scattering volumes.
Working at generic momentum and scattering angle, we express the corresponding resource powers in terms of these effective-range-expansion (ERE) parameters.
For magic and non-local magic, this extends the low-energy $NN$ analyses of Refs.~\cite{Robin:2024oqc,Robin:2025ymq} beyond the $S$ wave.

The remainder of the paper is organized as follows. In Sec.~\ref{sec:qi_measures} we set up the QI framework and define the resources used in our analysis.
In Sec.~\ref{sec:scattering} we discuss the relevant \eftnopi\ background and construct the $NN$ scattering amplitude in the partial-wave basis through NLO.
Section~\ref{sec:wigner} derives the operator structure implied by Wigner symmetry beyond the $S$ wave.
Section~\ref{sec:results} presents the physical and Wigner-symmetric results for the different resources and explains the distinct behavior of the $np$ and $nn$ channels.
The appendix contains additional details concerning the operators that enter the \eftnopi\ amplitude.

\section{Quantum information and scattering theory}
\label{sec:qi_measures}

\subsection{Scattering and spin state}
\label{sec:2a}

We consider elastic $NN$ scattering in the center-of-mass frame. Suppressing momentum labels, the two-nucleon spin Hilbert space is
\begin{equation}
\calH_{\rm spin} = \calH_{n}\otimes\calH_{p} \simeq \mathbb{C}^{2}\otimes\mathbb{C}^{2},
\end{equation}
spanned by the product basis $\{|\!\uparrow\uparrow\rangle,|\!\uparrow\downarrow\rangle,|\!\downarrow\uparrow\rangle,|\!\downarrow\downarrow\rangle\}$ of spin projections along a chosen quantization axis.
For fixed relative momentum $p$ and scattering direction $\hat{\vb{p}}'$, the spin-dependent scattering operator has the standard form
\begin{equation}\label{eq:Sop}
S(\hat{\vb{p}}',\hat{\vb{p}};p) = \1 + i\frac{M_{N}p}{2\pi}\,\calM(\hat{\vb{p}}',\hat{\vb{p}};p),
\end{equation}
where $\calM$ is the spin-space transition amplitude constructed from partial waves in Sec.~\ref{sec:scattering}. 
By rotational invariance, $\calM$ depends only on $p$ and the scattering angle $\Theta$ between $\hat{\vb{p}}$ and $\hat{\vb{p}}'$. 
The identity gate in Eq.~\eqref{eq:Sop} contributes only to the case of forward scattering, namely when $\Theta=0$.
The operator used to generate the plotted outgoing spin state is
\begin{equation}\label{eq:Toperator}
\calT(p,\Theta)=
\begin{cases}
S(\hat{\vb p},\hat{\vb p};p), & \Theta=0\,,\\
\calM(\hat{\vb p}',\hat{\vb p};p), & \Theta\ne0\,.
\end{cases}
\end{equation}
Thus the forward curves retain the contribution from the identity gate and use the full unitary $S$ operator, whereas the nonforward curves use the transition amplitude.

We prepare a separable initial spin state,
\begin{equation}
|\chi_{\rm in}\rangle = |\alpha\rangle\otimes|\beta\rangle,
\end{equation}
with $|\alpha\rangle,|\beta\rangle\in\mathbb{C}^{2}$.
At a nonforward scattering direction ($\hat{\vb{p}}'\neq\hat{\vb{p}}$), the transition amplitude is not unitary, so its action must be normalized. We therefore define
\begin{equation}\label{eq:chiout}
|\chi_{\rm out}\rangle = \frac{\calT\,|\chi_{\rm in}\rangle}{\sqrt{\langle\chi_{\rm in}|\calT^{\dagger}\calT|\chi_{\rm in}\rangle}},
\end{equation}
and the full density matrix of the two-nucleon spin system is
\begin{equation}\label{eq:rhochi}
\rho_{\rm out} = |\chi_{\rm out}\rangle\langle\chi_{\rm out}|.
\end{equation}
Since Eq.~\eqref{eq:chiout} defines a pure state at fixed kinematics, all entanglement measures below are computed on this pure bipartite state.

For the forward case in Eq.~\eqref{eq:Toperator}, unitarity makes the denominator in Eq.~\eqref{eq:chiout} equal to one. In the $S$-wave approximation this is the setting of Ref.~\cite{Beane:2018oxh}; tensor-induced $S$--$D$ mixing lies beyond that restriction and is omitted here.
This distinction resolves the seemingly contradictory observations in the literature: the entanglement suppression in Ref.~\cite{Beane:2018oxh} concerns the full $S$ operator, whereas the entanglement enhancement in parts of phase space in Refs.~\cite{Miller:2023ujx,Bai:2023tey} concerns normalized transition amplitudes in nonforward directions.

\subsection{Entanglement measures}

The bipartite entanglement between the neutron and proton spin subsystems is quantified by tracing out one particle. The reduced density matrices are
\begin{equation}
\rho_{n} = \mathrm{Tr}_{p}\,\rho_{\rm out},
\end{equation}
and $\rho_{p}=\mathrm{Tr}_{n}\,\rho_{\rm out}$. For the pure bipartite state in Eq.~\eqref{eq:rhochi}, $\rho_n$ and $\rho_p$ have the same nonzero eigenvalues, although they need not be equal as operators. The entanglement entropy is the von Neumann entropy of either reduced state,
\begin{equation}
S_{E}(\rho) = -\mathrm{Tr}\!\left(\rho_{n}\ln\rho_{n}\right).
\end{equation}
For separable states $S_{E}=0$, while for maximally entangled two-qubit states $S_{E}=\ln 2$.

In practice, it is useful to employ the linear entropy,
\begin{equation}
E_{\rm lin}(\rho) = 2\!\left(1-\mathrm{Tr}\,\rho_{n}^{2}\right),
\end{equation}
which corresponds to the leading expansion of the entanglement entropy around purity and ranges from $0$ (separable) to $1$ (maximally entangled) for two-qubit states.

\subsection{Entanglement power}

The entanglement measures introduced above depend on the choice of initial spin state $|\chi_{\rm in}\rangle$. To characterize the intrinsic entangling capability of the scattering dynamics, independent of any particular state preparation, we define the entanglement power~\cite{Zanardi:2000zz,Beane:2018oxh} as the average entanglement generated from all initially separable spin states.

A general single-nucleon spin state can be parametrized on the Bloch sphere,
\begin{equation}
|\phi\rangle = \cos\frac{\theta}{2}\,|\!\uparrow\rangle + e^{i\varphi}\sin\frac{\theta}{2}\,|\!\downarrow\rangle,
\end{equation}
with the Haar measure $d\mu(\phi) = \frac{1}{4\pi}\sin\theta\,d\theta\,d\varphi$. The entanglement power at fixed kinematics is then
\begin{equation}\label{eq:EP}
\calE(p,\hat{\vb{p}}') = \int d\mu(\alpha)\,d\mu(\beta)\;E_{\rm lin}\!\big(\rho_{\rm out}
)\,,
\end{equation}
where $\rho_{\rm out}$ is defined in Eq.~(\ref{eq:rhochi}) and depends on $|\chi_{\rm in}\rangle = |\alpha\rangle\otimes|\beta\rangle$.
This state-averaged quantity isolates the amount of spin entanglement that neutron--proton scattering typically generates when acting on unentangled incoming states, and can be thought of as a property of the scattering operator itself rather than of a particular initial configuration.

\subsection{Magic and non-local magic powers}

We also use the same normalized outgoing state in Eq.~\eqref{eq:chiout} to quantify magic, or nonstabilizerness.
Stabilizer states are generated from computational-basis states by Clifford operations and define the zero-magic sector used below~\cite{Nielsen:2012yss,Gottesman:1998hu,Aaronson:2004xuh,Emerson:2013zse}.
In addition, magic captures the non-Clifford resource needed for universal quantum computation~\cite{Bravyi:2004isx}.
For the explicit measure we use the stabilizer R\'enyi entropy (SRE) introduced in Ref.~\cite{Leone:2021rzd} and generalized to qudits in Ref.~\cite{Wang:2023uog}; see also Ref.~\cite{Ohta:2025utz} for recent work on extremal magic states.
For a pure two-qubit state $\rho=|\chi\rangle\langle\chi|$, the second stabilizer R\'enyi entropy is
\begin{equation}\label{eq:M2_state}
M_{2}(\rho)
=-\log\!\left[\frac{1}{4}\sum_{P\in\mathcal P_{2}}
\left(\mathrm{Tr}\,\rho P\right)^{4}\right],
\end{equation}
where $\mathcal P_{2}$ is the two-qubit Pauli group modulo phases.
This quantity vanishes for stabilizer states.
For two qubits its maximum is $M_{2}^{\rm max}=\log(16/7)$~\cite{Liu:2025frx}.

The SRE defined in Eq.~(\ref{eq:M2_state}) depends on choosing an  axis for spin quantization. Two popular choices are 1) the spin basis which projects the outgoing spins along the incoming beams \cite{Liu:2025qfl} and 2) the helicity basis which projects the spin along the direction of the motion of the outgoing particles \cite{White:2024nuc}. It is possible to define a ``non-local'' magic, which
minimizes a magic monotone over local unitaries acting separately on the two subsystems~\cite{Qian:2025oit,Gargalionis:2026onv,Busoni:2026lvp}.
For pure two-qubit states this minimization can be performed analytically, so the non-local magic used in Ref.~\cite{Robin:2025ymq} can be written directly in terms of the linear entropy,
\begin{align}\label{eq:MNL_state}
M_{\rm NL}(\rho)
&= \min_{U_n,U_p}M_2\!\left[
(U_n\otimes U_p)\rho
(U_n^\dagger\otimes U_p^\dagger)\right] \nonumber\\
&=-\log\!\left(1-E_{\rm lin}(\rho)+E_{\rm lin}^{2}(\rho)\right).
\end{align}
For the outgoing particles the non-local magic is invariant under the spatial rotations on the two qubits and, therefore, independent of the spin-quantization axis. It has been observed in Ref.~\cite{Gargalionis:2026onv} that in some scattering processes the non-local magic agrees with the magic computed in the helicity basis.
In the present work we compute magic in the spin basis and also calculate the non-local magic.
Eq.~\eqref{eq:MNL_state} vanishes for both separable states and  maximally entangled two-qubit states, which are locally equivalent to Bell stabilizer states, and reaches its maximum $\log(4/3)$ at $E_{\rm lin}=1/2$~\cite{Qian:2025oit,Robin:2025ymq}.
We therefore normalize the plotted non-local magic by $\log(4/3)$.

Similar to the entanglement power, the magic power characterizes the ability of scattering dynamics to generate magic from zero-magic inputs by averaging over initial stabilizer states. It has been used to study both high-energy collisions and low-energy nuclear physics~\cite{White:2024nuc,Liu:2025qfl,Gargalionis:2025iqs,Robin:2024oqc,Robin:2025ymq}.
Following the convention of Refs.~\cite{Robin:2024oqc,Robin:2025ymq}, we use the full two-qubit stabilizer ensemble, rather than only the product subset of one-qubit stabilizer states.
Let $\mathcal S_{2}$ denote the 60 pure stabilizer states of a two-qubit Hilbert space, equivalently the Clifford orbit of the computational-basis states.
This choice should be distinguished from a product-stabilizer average over $\mathcal S_{1}\times \mathcal S_{1}$, with $|\mathcal S_{1}|=6$, which is a useful product-input diagnostic but is not the same magic-power resource.
Because the normalized transition-amplitude resources at nonforward angles are not polynomial functions of the input state, we evaluate these finite stabilizer sums directly and do not replace them by Haar integrals using projective-design identities.
At fixed kinematics we define
\begin{align}
\mathcal M_{2}(p,\hat{\vb p}')
&=\frac{1}{|\mathcal S_{2}|}
\sum_{\sigma\in\mathcal S_{2}}
\frac{M_{2}\!\left(\rho[\calT\,|\sigma\rangle]\right)}{\log(16/7)}\,,\label{eq:MP}\\
\mathcal M_{\rm NL}(p,\hat{\vb p}')
&=\frac{1}{|\mathcal S_{2}|}
\sum_{\sigma\in\mathcal S_{2}}
\frac{M_{\rm NL}\!\left(\rho[\calT\,|\sigma\rangle]\right)}{\log(4/3)}\,.\label{eq:MNLP}
\end{align}

\section{Scattering amplitudes in effective field theory}
\label{sec:scattering}

We consider elastic neutron--proton scattering in the center-of-mass frame, with incoming and outgoing relative momenta $\vb p$ and $\vb p'$ satisfying $|\vb p'|=|\vb p|\equiv p$. 
The spin-space amplitude $\calM$ in Eq.~\eqref{eq:Sop} is expanded in partial waves of definite orbital angular momentum $\ell$, spin $s$, and total angular momentum $j$, each carrying a scalar amplitude $\calA^{(c)}$ for the channel $c\equiv{}^{2s+1}\ell_{j}$.
For an uncoupled partial wave $c$, the $S$-matrix element is
    \begin{equation}
        \bra{p;\ell s,jm_{j}}S\ket{p;\ell s,jm_{j}} = e^{2i\delta_{c}(p)},
    \end{equation}
with phase shift $\delta_{c}(p)$.
With the standard nonrelativistic normalization (see, e.g. \cite{Kaplan:1996xu, Chen:1999tn, vanKolck:1998bw}), the partial wave amplitude $\calA^{(c)}$ is related to the $S$-matrix according to
    \begin{align}
        e^{2 i \delta_c(p)} & = 1 + i \frac{M_N p}{2 \pi} \calA^{(c)}(p) \, ,
    \end{align}
where $M_N$ is the nucleon mass.
This can be rearranged into the form,
    \begin{equation}
    \label{eq:pw_amp}
        \calA^{(c)}(p) = \frac{4\pi}{M_{N}}\frac{1}{p\cot\delta_{c}(p)-ip}\,.
    \end{equation}
In the present work we neglect the $^{3}S_{1}$--$^{3}D_{1}$ tensor-force mixing, which enters at order $p^{2}$ relative to the leading $S$-wave amplitude and whose quantitative effects on entanglement have been discussed in Refs.~\cite{Bai:2023tey,Miller:2023ujx}.
Although this mixing enters at the same nominal order as the terms retained below, we isolate the leading angular dependence from the uncoupled $P$ waves; including coupled channels is left for future work.

For low momenta, the function $p \cot \delta_c(p)$ has the effective range expansion \cite{Bethe:1949yr}
    \begin{equation}\label{eq:ERE_Swave}
        p\cot\delta_{c}(p) = -\frac{1}{a^{(c)}} + \frac{1}{2}\,r_0^{(c)}\,p^{2} + \calO(p^{4}),
        \quad
        c\in\{{}^{1}S_{0},\,{}^{3}S_{1}\}\,,
    \end{equation}
where $a^{(c)}$ is the scattering length and $r_0^{(c)}$ is the effective range.
$NN$ scattering at low energies is characterized by unnaturally large scattering lengths in both channels ($a^{({}^{1}S_{0})}\simeq -23.7$~fm, $a^{({}^{3}S_{1})}\simeq 5.4$~fm)~\cite{Kaplan:1998we}, both well above the natural scale $m_{\pi}^{-1}\simeq 1.4$~fm set by the pion mass.
The effective ranges $r_0^{({}^1S_0)} \approx 2.7$ fm and $r_0^{({}^3S_1)} \approx 1.7$ fm, on the other hand, are considered natural, i.e., $r_0 \sim 1/m_\pi$.
For momenta $\left| \vb p \right| \sim 1/a^{(c)} \ll m_\pi$, the amplitude can be further expanded as,
    \begin{align}
        \label{eq:Swave_amp_expanded}
        \calA^{(c)} & = - \frac{4 \pi}{M_N} \frac{1}{\frac{1}{a^{(c)}} + i p} \left[1 + \frac{r_0^{(c)} p^2/2}{\frac{1}{a^{(c)}} + i p} + \cdots \right]\,,
    \end{align}
where the dots stand for higher order corrections.

This physics can also be captured in \eftnopi~\cite{Kaplan:1998we, Kaplan:1996xu, Kaplan:1998tg, vanKolck:1998bw}.
In particular, for the range of momenta we consider the LO contributions must be summed to all orders to reproduce the $p$ dependence of the amplitude in Eq.~\eqref{eq:Swave_amp_expanded}, while the NLO contributions are treated perturbatively (the complete Lagrangian through two-derivative operators is given in Appendix~\ref{app:operators}). 
The ${}^1S_0$ and ${}^3S_1$ amplitudes through NLO are then given by \cite{Kaplan:1998we}
    \begin{align}
        \calA^{(c)}(p) & = \calA^{(c)}_{-1}(p) + \calA^{(c)}_0(p)\,,  \label{eq:Swave_amp} \\
        \calA^{(c)}_{-1}(p) & = \frac{ -C_0^{(c)} }{1 + i C_0^{(c)} I_0(p)}\,, \label{eq:Swave_LO} \\
        \calA^{(c)}_0(p) & = \frac{-C^{(c)}_2 p^2}{\left(1 + i C_0^{(c)} I_0(p) \right)^2}\,, \label{eq:Swave_NLO}
    \end{align}
for $c\in\{{}^{1}S_{0},\,{}^{3}S_{1}\}$.
The two-nucleon bubble integral is given by
    \begin{equation}
        \label{eq:bubble_integral}
        I_0(p) = i M_N \int \frac{d^3 q}{(2 \pi)^3} \frac{1}{p^2 - q^2 + i \epsilon}\,.
    \end{equation}
This integral is power-law divergent, which requires the specification of a renormalization scheme.
Usually in \eftnopi, the power divergence subtraction (PDS) scheme \cite{Kaplan:1998we, Kaplan:1998tg} is used to determine the proper power counting.
In this scheme, the integral in Eq.~\eqref{eq:bubble_integral} takes the form \cite{Kaplan:1998we, Kaplan:1998tg}
    \begin{equation}
        I_0^{\rm PDS}(p, \mu) = \frac{M_N}{4 \pi} \left( p - i \mu \right)\,.
    \end{equation}
The $\mu$ dependence of the LECs is determined by the requirement that the amplitude is $\mu$ independent order-by-order and matching to Eq.~\eqref{eq:Swave_amp_expanded},
    \begin{align}
        C_0^{(c)}(\mu) & = \frac{4 \pi}{M_N} \frac{1}{\frac{1}{a^{(c)}} - \mu}\,, \\
        C_2^{(c)}(\mu) & = \frac{2 \pi}{M_N} \frac{r_0^{(c)}}{\left( \frac{1}{a^{(c)}} - \mu \right)^2}\,.
    \end{align}
With this setup, the amplitude is manifestly $\mu$ independent at each order as promised.
Furthermore, the results one would obtain in the minimal subtraction scheme are equivalent to those from PDS by taking $\mu = 0$.
For our purposes, the latter choice makes the connection between the quantum resources considered here and the physical scattering parameters more transparent.
Therefore, we will consider $\mu = 0$ in the remainder of this work.

The $P$-wave channels can be treated perturbatively in the regime we are considering.
In particular, we only retain the lowest order tree-level amplitude \cite{Schindler:2018irz}
    \begin{align}
        \calA^{\,(^1P_1)} & = \frac{1}{3} p^2 C_2^{\, (^1P_1)}\,, \label{eq:1P1_amp} \\
        \calA^{\,({}^3P_0)} & = p^2 C_2^{\,({}^3P_0)}\,, \label{eq:3P0_amp} \\
        \calA^{\,({}^3P_1)} & = \frac{2}{3} p^2 C_2^{\,({}^3P_1)}\,, \label{eq:3P1_amp} \\
        \calA^{\,({}^3P_2)} & = \frac{4}{3} p^2 C_2^{\,({}^3P_2)}\,. \label{eq:3P2_amp}
    \end{align}
In this case, the couplings are matched to the lowest order of the $P$-wave effective range expansion,
    \begin{equation}\label{eq:ERE_Pwave}
    \begin{gathered}
        p^{3}\cot\delta_{c}(p) = -\frac{1}{a^{(c)}} + \calO(p^{2}),\\
        c\in\{{}^{1}P_{1},\,{}^{3}P_{0},\,{}^{3}P_{1},\,{}^{3}P_{2}\}\,.
    \end{gathered}
    \end{equation}
This translates into the couplings as
    \begin{align}
        C_2^{\,({}^1P_1)} & = - 3 \frac{4 \pi a^{\,({}^1P_1)}}{M_N}\,, \\
        C_2^{\,({}^3P_0)} & = - \frac{4 \pi a^{\,({}^3P_0)}}{M_N}\,, \\
        C_2^{\,({}^3P_1)} & = - \frac{3}{2} \frac{4 \pi a^{\,({}^3P_1)}}{M_N}\,, \\
        C_2^{\,({}^3P_2)} & = - \frac{3}{4} \frac{4 \pi a^{\,({}^3P_2)}}{M_N}\,.
    \end{align}

The spin-space amplitude $\calM_{m_3 m_4, m_1 m_2}(\hat p', \hat p; p) = \bra{\vb p' m_3 m_4} \calM \ket{\vb p m_1 m_2}$ can be related to the partial wave amplitudes through,
\begin{align}\label{eq:Mfull}
&\calM_{m_{3}m_{4},m_{1}m_{2}}(\hat{\vb p}',\hat{\vb p};p) = 4\pi\!\sum_{\ell,s,j,m_{j}}\sum_{m_{\ell},m_{\ell}'}\sum_{m_{s},m_{s}'}\nonumber\\
&\quad\times Y_{\ell}^{m_{\ell}'}(\hat{\vb p}')\,Y_{\ell}^{m_{\ell}*}(\hat{\vb p})\;
\braket{\tfrac{1}{2}m_{3}\tfrac{1}{2}m_{4}}{sm_{s}'}\nonumber\\
&\quad\times\braket{\tfrac{1}{2}m_{1}\tfrac{1}{2}m_{2}}{sm_{s}}\nonumber\\
&\quad\times
\braket{\ell m_{\ell}'sm_{s}'}{jm_{j}}
\braket{\ell m_{\ell}sm_{s}}{jm_{j}}\;
\calA^{({}^{2s+1}\ell_{j})}(p)\,,
\end{align}
where $\ell$, $s$, $j$, and $m_j$ are the orbital angular momentum, total spin, total angular momentum, and total-angular-momentum projection quantum numbers, respectively.
Similarly, $m_\ell$ ($m_\ell'$) and $m_s$ ($m_s'$) denote the orbital and spin projections in the incoming (outgoing) partial-wave states.
The bra-kets in Eq.~\eqref{eq:Mfull} are standard Clebsch-Gordan coefficients.
$Y_\ell^{m_\ell}(\hat p)$ denotes the spherical harmonic of rank $\ell$, where $\hat p = \vb p/\abs{p}$ is the direction of the three-momentum $\vb p$.
The $4\times4$ transition amplitude $\calM$ entering Eq.~\eqref{eq:Toperator} is thus fully determined by the set of partial-wave amplitudes $\{\calA^{(c)}\}$ through Eq.~\eqref{eq:Mfull}.

\section{Entanglement in and away from the Wigner limit}
\label{sec:wigner}

At very low energy, the $np$ amplitude is dominated by the $S$-wave contribution and the spin-space amplitude reduces to a linear combination of the identity and SWAP quantum gates~\cite{Low:2021ufv},
\begin{eqnarray}\label{eq:Mswave}
\calM_{np}^{(S)}\big|_{p\to 0} &=& -\frac{2\pi}{M_{N}}\!\left[(a^{({}^{3}S_{1})}+a^{({}^{1}S_{0})})\,\1_s \right.\nonumber\\
 &&\ \ +\left. (a^{({}^{3}S_{1})}-a^{({}^{1}S_{0})})\,P_{\rm SWAP}\right],
\end{eqnarray}
where $\1_s$ is the identity gate in spin space and $P_{\rm SWAP}=\frac{1}{2}(\1_s+\boldsymbol{\sigma}_{n}\cdot\boldsymbol{\sigma}_{p})$ is the SWAP gate, which exchanges the nucleon spins.
Entanglement is generated only by the $P_{\rm SWAP}$ term; the entanglement power therefore vanishes when $a^{({}^{1}S_{0})}=a^{({}^{3}S_{1})}$, which is precisely the Wigner $SU(4)$ limit~\cite{Beane:2018oxh,Low:2021ufv}. 
The anomalously large and \emph{unequal} physical scattering lengths ($|a_{{}^{1}S_{0}}|\gg a_{{}^{3}S_{1}}$) thus encode both the proximity to unitarity and the degree of Wigner-symmetry breaking, making the $S$-wave sector the primary driver of entanglement at low momenta.

At higher momenta, the two-derivative operators in the $S$- and $P$-waves can also be constrained by Wigner symmetry.
In the framework of \eftnopi, the role of Wigner symmetry is best seen through a change of basis.
At zero-derivative order, Wigner symmetry implies that $C_0^{\,({}^1S_0)} = C_0^{\,({}^3S_1)}$ \cite{Mehen:1999qs}.
At two-derivative order, there is a total of 14 possible contact operators invariant under parity, rotations, Galilean boosts, and isospin which can be reduced to a complete set of 7 possible operators \cite{Ordonez:1995rz, Epelbaum:1998ka, Girlanda:2010ya, Schindler:2018irz}.
To analyze the Wigner symmetric interaction, the most convenient basis is that of Ref.~\cite{Schindler:2018irz}.
Specifically, the part of the Lagrangian that could preserve Wigner symmetry is given by
    \begin{equation}
    \label{eq:wigner_two_derivative}
    \begin{split}
        \mathcal L & \supset C_{1 \cdot 1} \nabla^i \left( N^\dagger N \right) \nabla^i \left( N^\dagger N \right)  \\
        & + C_{\sigma \cdot \sigma} \nabla^i \left( N^\dagger \sigma^j N \right) \nabla^i \left( N^\dagger \sigma^j N \right) \\ 
        & + C_{\tau \cdot \tau} \nabla^i \left( N^\dagger \tau^a N \right) \nabla^i \left( N^\dagger \tau^a N \right) \\
        & + C_{G \cdot G} \nabla^i \left( N^\dagger \sigma^j \tau^a N \right) \nabla^i \left( N^\dagger \sigma^j \tau^a N \right)\,.
    \end{split}
    \end{equation}
The Wigner-symmetric limit is then given by
    \begin{equation}\label{eq:wigner}
        C_{\sigma \cdot \sigma} = C_{\tau \cdot \tau} = C_{G \cdot G}\,,
    \end{equation}
in which case the resulting operator is proportional to the Wigner ${\rm SU}(4)$ quadratic Casimir \cite{Wigner:1936dx, Wigner:1939zz, Hecht:1969ck, LiMuli:2025zro} (see also Refs.~\cite{Phillips:2013rsa, Kaplan:1995yg} in the context of the large-$N_c$ expansion),
    \begin{equation}
        C_2 = \frac{1}{2} S^2 + \frac{1}{2} I^2 + 2 G^{ia} G^{ia}\,,
    \end{equation}
where $\vb S$ is the total spin operator, $\vb I$ is the total isospin operator, and $G^{ia}$ is the combined spin-isospin generator with spin (isospin) index $i$ ($a$).
In this case, the partial wave couplings become \cite{Schindler:2018irz}
    \begin{equation}
    \begin{split}
        C_2^{\,({}^1S_0)} & = C_2^{\,({}^3S_1)} = -4 \left( C_{1 \cdot 1} - 5 C_{G \cdot G} \right)\,, \\
        C_2^{\,({}^1P_1)} & = -4 \left( C_{1 \cdot 1} + 3 C_{G \cdot G} \right)\,, \\
        C_2^{\,({}^3P_0)} & = -\frac{4}{3} \left( C_{1 \cdot 1} + 3 C_{G \cdot G} \right)\,, \\
        C_2^{\,({}^3P_1)} & = -2 \left( C_{1 \cdot 1} + 3 C_{G \cdot G} \right)\,, \\
        C_2^{\,({}^3P_2)} & = - \left( C_{1 \cdot 1} + 3 C_{G \cdot G} \right)\,.
    \end{split}
    \label{eq:wigner_couplings}
    \end{equation}
As a result, the effective ranges in the ${}^1S_0$ and ${}^3S_1$ channels are equal and the $P$-wave scattering volumes are also equal.
The remaining three two-derivative operators break Wigner symmetry explicitly; in particular they contain the spin-orbit force and the tensor force.

In the Wigner limit of the neutron--proton channel, the relations between the couplings imply $\calA^{(\oneS)} = \calA^{(\threeS)} \equiv \calA_S$ and $\calA^{(\onePone)} = \calA^{(\PJ)} \equiv \calA_P$ for $J = 0$, $1$, and $2$.
Succinctly, the scattering amplitude is independent of the total spin $s$ of the system.
Completeness of the Clebsch-Gordan coefficients and the spherical harmonic addition theorem then reduce Eq.~\eqref{eq:Mfull} to
    \begin{align}
        & (\calM_{np})_{m_{3}m_{4},m_{1}m_{2}}(\hat{\vb p}',\hat{\vb p};p) = \sum_{l,s,m_s}  \calA_l(p) \nonumber \\
        & \times (2l + 1) P_l(\hat p' \cdot \hat p) 
        \braket{\tfrac{1}{2}m_{3}\tfrac{1}{2}m_{4}}{sm_{s}}
        \braket{\tfrac{1}{2}m_{1}\tfrac{1}{2}m_{2}}{sm_{s}}\,.
    \end{align}
The sum over $s$ and $m_s$ then reduces each term in the $l$ expansion to the identity gate in spin space,
    \begin{align} \label{eq:M_wig_np}
        \calM_{np}(\hat{\vb p}',\hat{\vb p};p) \simeq
        \left(\calA_S(p) + 3 \calA_P(p) P_1(\hat p' \cdot \hat p)\right)\1_s\,.
    \end{align}

This is one of the central findings of this work: beyond the $S$ wave, Wigner's $SU(4)$ symmetry continues to force the outgoing state to factorize between spin and kinematics because the spin-space amplitude is proportional to the identity gate.
We highlight this particular finding because, generically, the outgoing state is a highly entangled state in the spin and momentum degrees of freedom: the spin orientation of the outgoing particles depends on their location in the outgoing phase space. Indeed, in Refs.~\cite{Bai:2023tey,Miller:2023ujx} the degree of entanglement in the scattered particles is a complicated function of the kinematics. It is intriguing that, in the Wigner-symmetric limit, the outgoing state suddenly collapses into an unentangled state between spin and momentum.

Now consider the same-nucleon channels.\footnote{If we neglect isospin-breaking effects from the quark masses and electromagnetism, $nn$ and $pp$ scattering are equivalent. For brevity, we refer to both channels in this limit as $nn$.}
The isospin component of the two-neutron wave function is clearly in a symmetric representation.
Therefore, the combined spatial and spin parts of the wave function must be antisymmetric in accord with Fermi statistics.
This forbids ${}^{3}S_{1}$ ($L=0,\,S=1$) and ${}^{1}P_{1}$ ($L=1,\,S=0$) channels and we cannot invoke the completeness of the Clebsch-Gordan coefficients in Eq.~\eqref{eq:Mfull}.
Therefore, while the Wigner limit implies the $np$ amplitude is proportional to $\1_s$, the same is not true in the $nn$ amplitude, which can be written as
\begin{align}
\calM_{nn}&\simeq
\calA^{(\oneS)}(p)\,P_A
 \notag \\
&+3\calA^{({}^{3}P_J)}(p)\,
P_1(\hat p' \cdot \hat p)\,
\left(\1_s-P_A\right)\,,
\label{eq:M_wig_nn}
\end{align}
where 
\begin{equation}
    P_A = |\chi_-\rangle\langle\chi_-| \,, \quad |\chi_-\rangle=\frac{1}{\sqrt{2}}\left(|\!\uparrow\downarrow\rangle-|\!\downarrow\uparrow\rangle\right)\,,
    \label{eq:PA}
\end{equation}
is the singlet projector that acts trivially on the one-dimensional antisymmetric spin Hilbert space, as discussed in Ref.~\cite{Hu:2025lua}.

\section{Results for quantum resources}
\label{sec:results}

\begin{figure*}[t]
    \centering
    \includegraphics[width=0.98\textwidth]{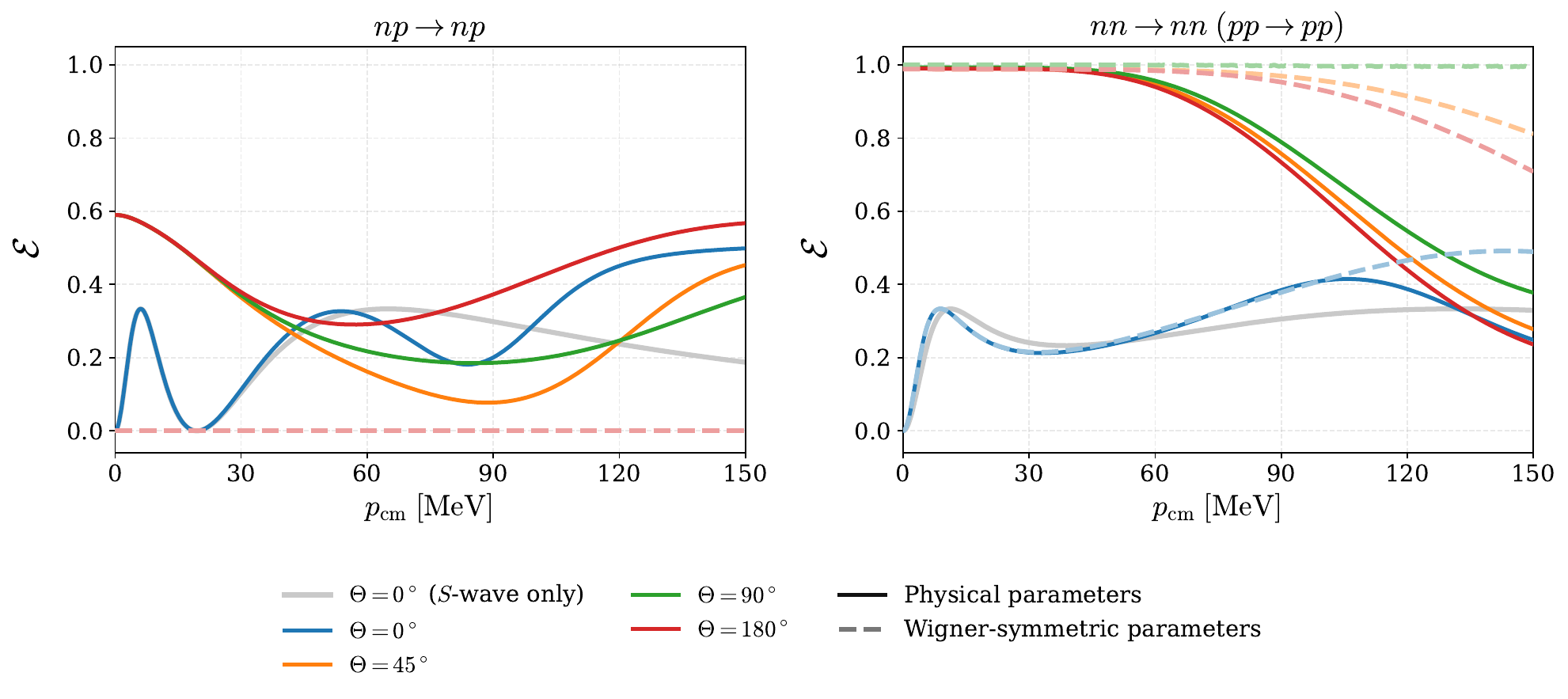}
    \caption{
        Entanglement power $\mathcal{E}$, Eq.~\eqref{eq:EP}, versus center-of-mass momentum for $\Theta=0^\circ,45^\circ,90^\circ,180^\circ$ in $np\to np$ (left) and $nn\to nn$ (right).
        Curves are colored by scattering angle.
        For each angle, colored solid lines use the physical ERE parameters and colored dashed lines use the Wigner-symmetric benchmark.
        The $\Theta=0^\circ$ curves use the full $S$ operator, while the nonforward curves use the normalized transition amplitude, as specified in Eq.~\eqref{eq:Toperator}.
        The solid light-gray curve is a forward reference using the physical ERE parameters and the full unitary $S$ operator restricted to the $S$-wave sector; in the right panel this is the Pauli-restricted same-nucleon analogue, where antisymmetry removes the triplet $S$ wave.
    }
    \label{fig:phys_vs_wigner}
\end{figure*}

\begin{table}[t]
\caption{
ERE inputs used for the solid curves in Fig.~\ref{fig:phys_vs_wigner}. These rounded low-energy $np$ parameters are taken from the NijmII analysis of Ref.~\cite{PavonValderrama:2005ku}; the $S$-wave scattering lengths and effective ranges are in fm, while the $P$-wave scattering volumes are in fm$^{3}$.
}

\label{tab:ere_inputs}
\begin{ruledtabular}
\begin{tabular}{lcc}
Channel & Input & Value  \\ 
\hline 
${}^{1}S_{0}$ & $a$, $r$ & $-23.7,\;2.7$ \\
${}^{3}S_{1}$ & $a$, $r$ & $5.42,\;1.753$ \\
${}^{1}P_{1}$ & $a^{(P)}$ & $2.80$ \\
${}^{3}P_{0}$ & $a^{(P)}$ & $-2.47$ \\
${}^{3}P_{1}$ & $a^{(P)}$ & $1.53$ \\
${}^{3}P_{2}$ & $a^{(P)}$ & $-0.28$ \\
\end{tabular}
\end{ruledtabular}
\end{table}

For reproducibility, the dashed $np$ benchmark sets both $S$-wave channels to $a=5.42$~fm and $r_0=1.753$~fm and all four $P$-wave scattering volumes to $a^{(P)}=0.1975$~fm$^3$.
The dashed $nn$ benchmark uses the common formal $S$-wave values $a=-23.7$~fm and $r_0=2.7$~fm and the common $P$-wave volume $a^{(P)}=-0.20333$~fm$^3$ before the Pauli-forbidden channels are projected out.

We now discuss the numerical results for the spin quantum resources generated by the physical and Wigner-symmetric amplitudes.
Fig.~\ref{fig:phys_vs_wigner} displays the entanglement power $\calE(p,\Theta)$ from Eq.~\eqref{eq:EP}, evaluated using the physical ERE parameters of Table~\ref{tab:ere_inputs} (solid curves) and in the Wigner-symmetric limit (dashed curves), for both $np\to np$ (left panel) and $nn\to nn$ (right panel) scattering at four representative scattering angles.
The solid light-gray curve isolates the forward reference obtained from the same physical ERE scattering lengths and effective ranges, but using the full unitary $S$ operator restricted to the $S$-wave sector.\footnote{For this unitary reference the linear-entropy integrand is a degree-two polynomial in the input density matrices. The product-Haar average can therefore be replaced exactly by an average over the one-qubit stabilizer 2-design, which explains the corresponding observation in Ref.~\cite{Robin:2024oqc}. This shortcut does not apply to the nonforward transition-amplitude resources in Eqs.~\eqref{eq:EP}, \eqref{eq:MP}, and \eqref{eq:MNLP}, where the outgoing state contains an input-dependent normalization.}
Ref.~\cite{Beane:2018oxh} studied the same $S$-wave entanglement-power resource using empirical phase shifts for the full unitary $S$ operator.
The gray curve shown here instead uses the ERE parameters of Table~\ref{tab:ere_inputs}, so agreement with phase-shift based results should only be expected within the ERE domain of validity.

In the Wigner-symmetric limit in the $np$ channel, the entanglement power is identically zero for any scattering angle.
With the physical scattering parameters, the forward-direction ($\Theta=0$) curve reduces at low momenta to the pure $S$-wave expression  $\calE=\tfrac{1}{3}\sin^{2}\!\big(2(\delta_{{}^{3}S_{1}}-\delta_{{}^{1}S_{0}})\big)$ in the linear-entropy normalization of Eq.~\eqref{eq:EP}, with the same phase-shift dependence as in Ref.~\cite{Beane:2018oxh}. 
The low-energy peak in $\calE$ occurs near the momentum where $|\delta_{{}^{3}S_{1}}-\delta_{{}^{1}S_{0}}|\simeq\pi/4$, at which point $\calE$ saturates its maximum value of $1/3$.
Then $\calE$ vanishes around $p \sim 20$ MeV where $|\delta_{{}^{3}S_{1}}-\delta_{{}^{1}S_{0}}| = \pi/2$.
Around $p \sim 60$ MeV, $\calE$ in the forward limit begins to receive stronger contributions from the $P$-wave amplitude leading to an overall upward trend compared to the decay seen in Ref.~\cite{Beane:2018oxh}.
Away from the forward limit, the entanglement power starts from a common finite low-momentum value, falls with angle-dependent structure, and rises again as $p$ approaches the breakdown scale of \eftnopi.

For nonforward angles in the Wigner-symmetric $nn$ sector, $\calE$ saturates at a $\Theta$-independent plateau at small $p$, where only the ${}^1S_0$ channel contributes significantly to the transition amplitude; the forward curve follows the full-$S$ convention of Eq.~\eqref{eq:Toperator}.
As shown in Eqs.~\eqref{eq:M_wig_nn} and \eqref{eq:PA}, this low-momentum amplitude is the singlet projector $P_A$; thus, any generic product-state inputs with nontrivial overlap with the spin-singlet state are mapped to a maximally entangled Bell state, which explains the $\calE\simeq1$ plateau.
As $p$ grows, $P$-wave contributions become more significant and generate the dependence on the scattering angle.
At very low momentum, the physical and Wigner-symmetric nonforward curves coincide because both reduce to the same singlet projection. They separate once the $P$ waves become relevant, with the physical curves lying below the Wigner-symmetric benchmark for $p\gtrsim50$~MeV.

Several features of the results in Fig.~\ref{fig:phys_vs_wigner} deserve comment, as they connect to a broader discussion in the recent literature. 
Ref.~\cite{Beane:2018oxh} emphasized that in the region $p\simeq 200$--$300$~MeV the entanglement power becomes small, and interpreted this as entanglement suppression tied to the approximate Wigner $SU(4)$ symmetry. 
However, the $S$-wave phase shifts vanish as $p \to \infty$ such that the $S$ matrix reduces to the identity gate.
In this particular energy region, the empirical ${}^1S_0$ and ${}^3S_1$ phase shifts from the Nijmegen database~\cite{Stoks:1993tb, PavonValderrama:2005ku} do have a downward trend; however, it is not immediately clear that this is due to symmetry rather than a general result from scattering theory.

Ref.~\cite{Miller:2023ujx} studied the same $np$ system but used a different entanglement resource: the outgoing two-nucleon state for a \emph{fixed} initial spin configuration $|\!\uparrow\downarrow\rangle$, rather than the entanglement power averaged over the two Bloch spheres, and also considered nonforward scattering. For this initial state, the outgoing-state entanglement is maximized at very low energies where the magnitude of the $S$-wave phase-shift difference passes through $\pi/4$, leading to the observation of ``entanglement maximization.''
This is the distinction anticipated in the Introduction: the two observations refer to different kinematic regimes and different entanglement measures.

Ref.~\cite{Bai:2023tey} generalized the analysis of Ref.~\cite{Beane:2018oxh} beyond the $S$-wave approximation, computing $\calE(p,\Theta)$ and related spin-state resources using the full phenomenological $S$-matrix at arbitrary $p$ and $\Theta$.
This revealed the significant angular dependence visible in the left panel of Fig.~\ref{fig:phys_vs_wigner}, arising from the $P$-wave channels that become important as $p$ grows.
The solid curves in the left panel of Fig.~\ref{fig:phys_vs_wigner} are consistent with the results of Ref.~\cite{Bai:2023tey} over the low-momentum region where the ERE is expected to apply.
Our ERE-based treatment additionally identifies how the individual scattering parameters enter the entanglement structure. Furthermore, we consider the Wigner-symmetric limit beyond $S$-wave scattering for both $np$ and $nn$ systems.

Ref.~\cite{Cavallin:2025kjn} performed a systematic study of both $np$ and $nn$ entanglement using potentials derived from chiral EFT with various levels of Wigner $SU(4)$ symmetry incorporated.
In line with our findings, Ref.~\cite{Cavallin:2025kjn} observed that entanglement suppression is \emph{not} present in the $nn$ channel and attributed this to the Pauli principle restricting the set of available partial waves.
Our analysis complements this observation by providing an explicit, amplitude-level explanation in terms of the completeness of the spin-coupled partial-wave basis, as discussed in Sec.~\ref{sec:wigner}.

\begin{figure*}[t]
    \centering
    \includegraphics[width=0.98\textwidth]{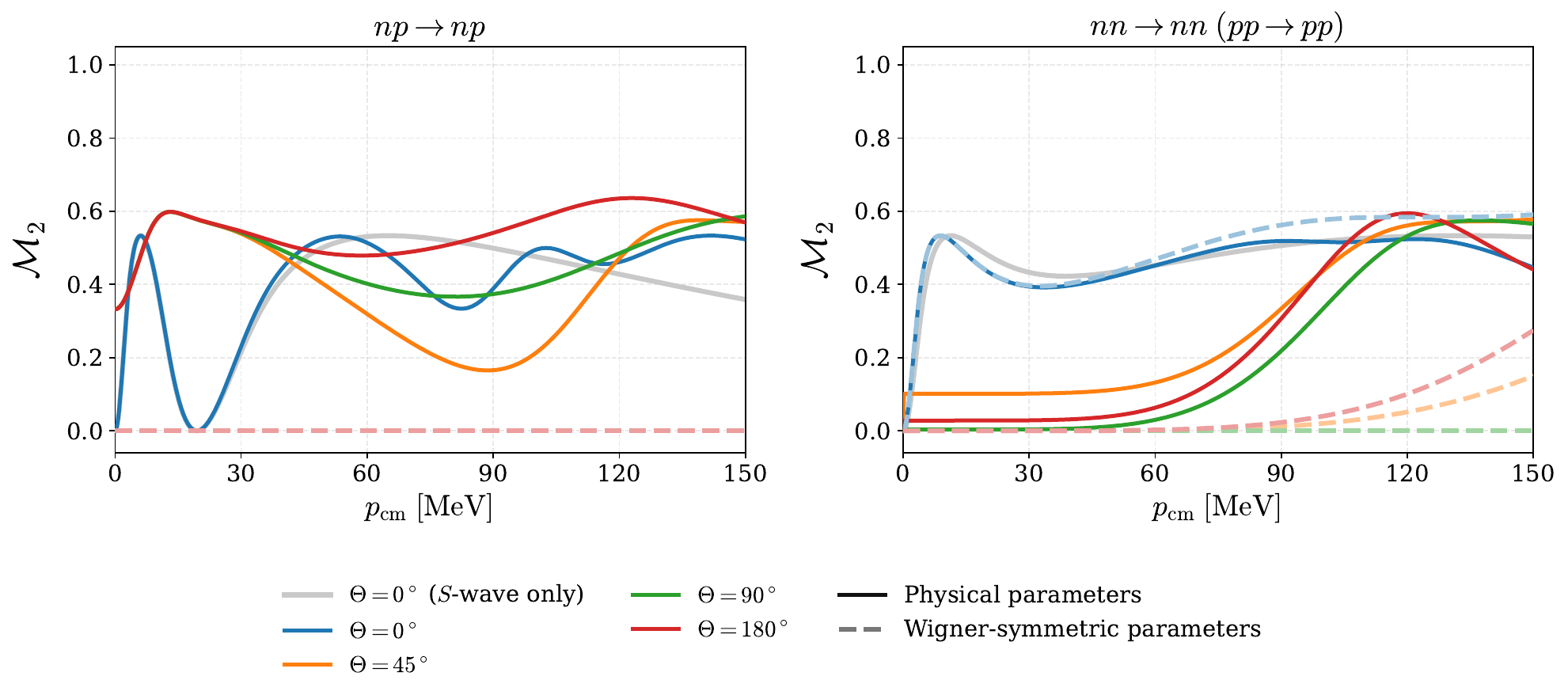}
    \caption{
        Same as Fig.~\ref{fig:phys_vs_wigner}, but for the two-qubit-stabilizer-averaged magic power $\mathcal M_{2}$, Eq.~\eqref{eq:MP}.
        The light-gray curve uses the same stabilizer-average convention as Ref.~\cite{Robin:2024oqc}.
    }
    \label{fig:phys_vs_wigner_magic}
\end{figure*}

\begin{figure*}[t]
    \centering
    \includegraphics[width=0.98\textwidth]{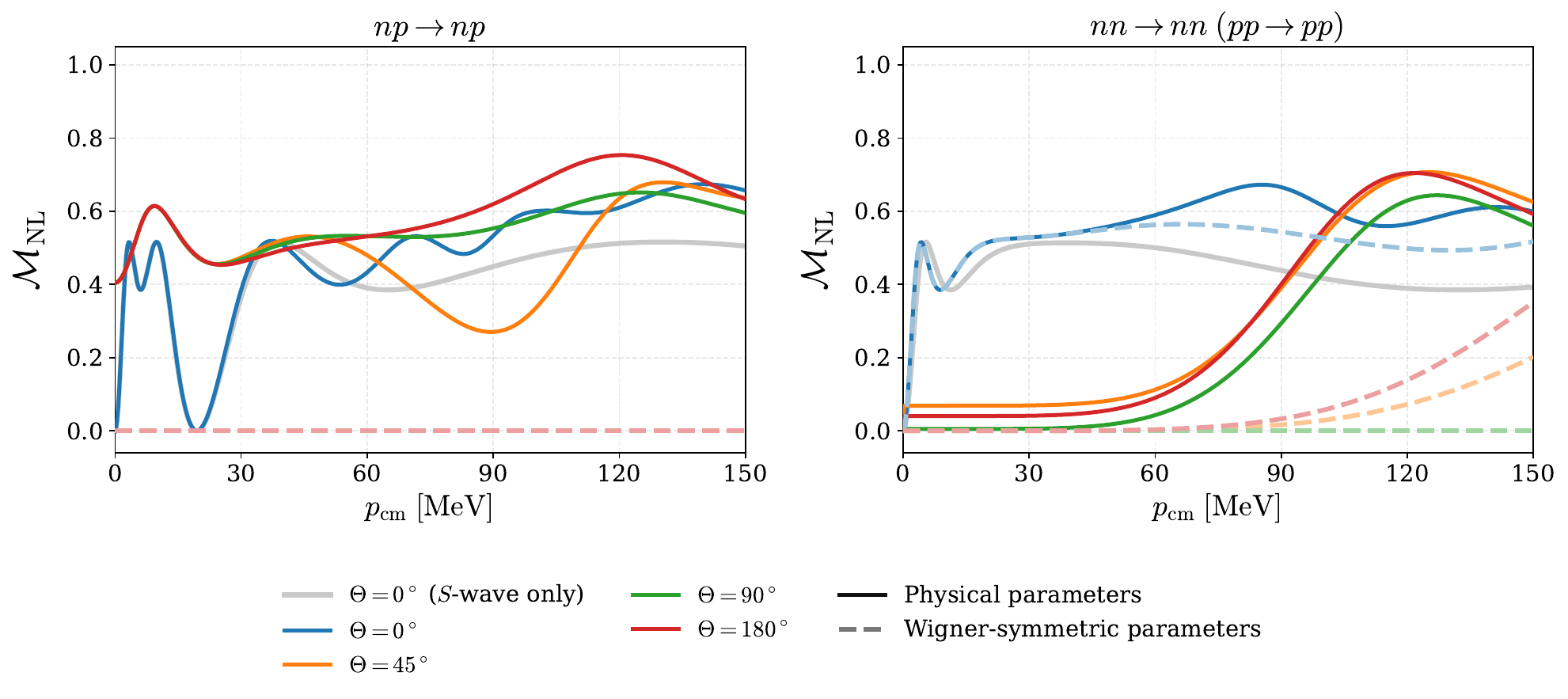}
    \caption{
        Same as Fig.~\ref{fig:phys_vs_wigner}, but for the two-qubit-stabilizer-averaged non-local magic power $\mathcal M_{\rm NL}$, Eq.~\eqref{eq:MNLP}.
        The light-gray curve uses the same stabilizer-average convention as Ref.~\cite{Robin:2025ymq}.
    }
    \label{fig:phys_vs_wigner_nlm}
\end{figure*}

Figs.~\ref{fig:phys_vs_wigner_magic} and~\ref{fig:phys_vs_wigner_nlm} show the corresponding magic and non-local magic powers computed from the same \eftnopi\ amplitudes.
The averages are taken over all 60 two-qubit stabilizer states, so these curves are directly comparable to the magic-power convention used in Refs.~\cite{Robin:2024oqc,Robin:2025ymq}.
The light-gray curves show the corresponding forward reference built from the unitary $S$ operator restricted to the $S$-wave sector.
In the Wigner-symmetric $np$ channel, both magic measures vanish for the same reason that the entanglement power vanishes: Eq.~\eqref{eq:M_wig_np} is proportional to the identity gate in spin space and therefore generates neither entanglement nor magic from initial states.
The physical $np$ curves are therefore a direct measure of Wigner-symmetry breaking in the spin amplitude. Their angular dependence is generated by the $P$ waves and is qualitatively similar to that of the entanglement power.

The identical-nucleon channel behaves differently, as expected. 
Again, in the Wigner-symmetric and low-energy $S$-wave limit, the operator $P_A$ in Eq.~\eqref{eq:PA} projects generic input states onto a Bell state $|\chi_-\rangle$.
Unlike entanglement, which is maximized for such a state, $|\chi_-\rangle$ is also a stabilizer state; thus, it carries no magic or non-local magic. 
At larger momenta, however, the amplitude is no longer a pure singlet projection and the triplet $P$-wave projection can generate magic from stabilizer inputs. In nonforward directions, both magic measures tend to be small where the entanglement is large and vice versa.
As follows from Eq.~\eqref{eq:MNL_state}, $\mathcal M_{\rm NL}$ is therefore a useful complement to $\calE$ in regions where the outgoing spin state is highly entangled but remains close to the stabilizer set.

Refs.~\cite{Robin:2024oqc,Robin:2025ymq} already established that low-energy $NN$ interactions generate magic and non-local magic in the $S$-wave regime.
For the full unitary $S$-wave reference, our formulas have the same phase-shift dependence: the resources depend only on $\delta_{{}^{3}S_{1}}-\delta_{{}^{1}S_{0}}$ and vanish in the Wigner-symmetric limit, with the stabilizer inputs grouped into the simple classes identified in those works. By contrast, the normalized transition amplitude depends separately on the two phase shifts through their relative channel weights.
The effective field theory reproduces this low-energy behavior by construction.
The new information in Figs.~\ref{fig:phys_vs_wigner_magic} and~\ref{fig:phys_vs_wigner_nlm} is the angular dependence introduced through the $P$-wave contributions.
These contributions split the simple stabilizer-input pattern, and show that total magic and non-local magic respond differently once higher partial waves are retained.
Consequently, $\mathcal M_{\rm NL}$ is not merely a rescaled entanglement curve, and its extrema can occur at intermediate entanglement.

\section{Conclusions}
\label{sec:conclusions}

We have extended the EFT description of spin quantum resources in low-energy $NN$ scattering by including higher-order operators and partial waves, connecting the resulting resource powers directly to scattering lengths, effective ranges, and scattering volumes.
This makes the deformation due to both momentum and angular dependence transparent.

Our results clarify four points in particular. First, we show how the $S$-wave sector controls the low-energy magnitude of the resource powers, and compute the nontrivial angular dependence emerging at higher energies via the $P$-wave scattering volumes in the EFT expansion. Second, the entanglement-suppression picture of Ref.~\cite{Beane:2018oxh} and the low-energy fixed-state enhancement emphasized in Ref.~\cite{Miller:2023ujx} probe different entanglement measures and different kinematic regimes. Third, the Wigner-symmetric case admits a simple operator interpretation.
In the $np$ channel, the Wigner limit beyond the $S$ wave continues to force the transition matrix to be proportional to the identity gate in spin space, so it does not generate new quantum resources upon scattering.
In the $nn$ channel, the Pauli principle removes the complementary ${}^{3}S_{1}$ and ${}^{1}P_{1}$ channels, so the argument that makes the Wigner-symmetric $np$ transition matrix proportional to the identity gate in spin space does not carry over to the $nn$ transition matrix in the interacting theory.
Fourth, the nonstabilizerness resources distinguish structures that are not visible in entanglement alone, because maximally entangled stabilizer states are not necessarily magic. To our knowledge, this is the first calculation of non-local magic in $nn$ scattering and, more generally, the first nonforward analysis of magic in $NN$ scattering.

This framework can be extended in several directions. The most immediate refinements are to include coupled channels such as ${}^{3}S_{1}$--${}^{3}D_{1}$. 
On the phenomenological side, it would also be interesting to connect the analytic parameter dependence found here to polarized spin measurements and to kinematic configurations that could maximize experimentally accessible spin entanglement in few-nucleon scattering. 
A further direction is to include interactions that violate fundamental symmetries such as $CP$ and study how the quantum-resource powers depend on their coefficients.

\begin{acknowledgments}
    We would like to thank Evgeny Epelbaum, John Gargalionis, David B. Kaplan, Matthias Schindler, Oliver Thim, and Jordy de Vries for valuable discussions.
    This work was supported by the NSF through cooperative agreement 2020275 (T.~R.~R.).
    T.~R.~R. thanks the Institute for Nuclear Theory at the University of Washington for its kind hospitality and stimulating research environment during the INT-26-1 program ``Nuclear Hamiltonians for Advancing Nuclear Physics and Beyond.''
    This research was supported in part by the INT's U.S. Department of Energy grant No. DE-FG02-00ER41132.
    This research used the CBorg AI platform and resources provided by the IT Division at the Lawrence Berkeley National Laboratory (Supported by the Director, Office of Science, Office of Basic Energy Sciences, of the U.S. Department of Energy under Contract No. DE-AC02-05CH11231). The initial phases of this work were supported by the Office of High Energy Physics of the U.S. Department of Energy (DOE) under Grant No. DE-SC0012567, and by the DOE QuantISED program through the theory consortium ``Intersections of QIS and Theoretical Particle Physics'' at Fermilab (FNAL 20-17). S.T. is supported by the Swiss National Science Foundation project number P5R5PT\_222350, and acknowledges CERN TH Department for hospitality while this research was being carried out. I.~L. is supported in part by the U.S. Department of Energy under contracts DE-AC02-06CH11357 (Argonne), DE-SC0023522 (Northwestern), DE-SC0010143 (Northwestern), and No. 89243024CSC000002 (QuantISED Program).
\end{acknowledgments}

\appendix

\section{NLO Contact Basis and Channel Matching}
\label{app:operators}

For completeness we collect here the contact-operator basis underlying the discussion in Sec.~\ref{sec:wigner}.
At zero derivatives, the most general two-nucleon contact Lagrangian can be written as~\cite{Weinberg:1990rz, Weinberg:1991um, Kaplan:1996xu, Kaplan:1998tg, Kaplan:1998we, vanKolck:1998bw}
\begin{equation}
\calL_{NN}^{(0)} =
-\frac{1}{2} C_S (N^\dagger N)(N^\dagger N)
-\frac{1}{2} C_T (N^\dagger \sigma_i N)(N^\dagger \sigma_i N),
\end{equation}
where $N=(p,n)^T$ is the nucleon doublet.
These couplings are related to those in partial wave basis according to
    \begin{align}
        C_0^{({}^1S_0)} & = C_S - 3 C_T\,, \\
        C_0^{({}^3S_1)} & = C_S + C_T\,.
    \end{align}
The operator proportional to $C_T$ breaks Wigner ${\rm SU}(4)$ explicitly.
Therefore, Wigner invariance is restored by requiring $C_T = 0$, which implies that the partial wave couplings are equal to one another.

The two-derivative contact operators have been derived in several places \cite{Ordonez:1995rz, Epelbaum:1998ka, Girlanda:2010ya, Schindler:2018irz}.
We find that the role of Wigner symmetry is more transparent in the basis retained in Ref.~\cite{Schindler:2018irz}; the following draws specifically from that work.
There are seven independent operators,
\begin{widetext}
\begin{align}
\label{eq:two_derivative_ops}
  \calL_{NN}^{(2)} & =  C_{1 \cdot 1} \nabla_i (N^\dagger N )  \nabla_i (N^\dagger N )  + C_{G \cdot G} \nabla_i (N^\dagger \sigma_j \tau_a N )  \nabla_i (N^\dagger \sigma_j \tau_a N  ) + C'_{G \cdot G} \nabla_i (N^\dagger \sigma_i \tau_a N )  \nabla_j (N^\dagger \sigma_j \tau_a N  ) \\
    &  
    + C_{\tau \cdot \tau} \nabla_i (N^\dagger \tau_a N )  \nabla_i (N^\dagger \tau_a N ) 
    + C_{\sigma \cdot \sigma} \nabla^i \left( N^\dagger \sigma^j N \right) \nabla^i \left( N^\dagger \sigma^j N \right)
    + \couplingSixZero \nabla_i (N^\dagger \sigma_i  N )  \nabla_j (N^\dagger \sigma_j  N  ) \nonumber \\
    & - \frac{i}{2}\couplingFiveZero \ \epsilon_{ijk} \left[ \nabla_j (N^\dagger \sigma_i  N )  (N^\dagger \LRd_k  N) + \nabla_j (N^\dagger  N )  (N^\dagger \LRd_k \sigma_i N  ) \right]\,.
\end{align}
The operators proportional to $C_{1 \cdot 1}$, $C_{G \cdot G}$, $C_{\sigma \cdot \sigma}$, and $C_{\tau \cdot \tau}$ were presented already in Eq.~\eqref{eq:wigner_two_derivative}.
The terms proportional to $\couplingFiveZero$, $C_{G \cdot G}'$, and $C_{\sigma \cdot \sigma}'$ break Wigner ${\rm SU}(4)$ explicitly.
Therefore, the Wigner limit is given by $\couplingFiveZero=C_{G \cdot G}'=C_{\sigma \cdot \sigma}'=0$ with $C_{G \cdot G}=C_{\sigma \cdot \sigma}=C_{\tau \cdot \tau}$ as discussed in Sec.~\ref{sec:wigner}.
Through a set of Fierz transformations, these operators can be rearranged in a way that makes their partial wave content manifest,
\begin{align}
\label{eq:two_derivative_partial_wave}
  \calL_2 & = \frac{1}{8} C_2^{(\threeS)} \left[ (N^T P_i N)^\dagger (N^T P_i \LRd^2 N) + {\rm H.c.} \right] + \frac{1}{8} C_2^{(\oneS)} \left[ (N^T P_a N)^\dagger (N^T P_a \LRd^2 N) + {\rm H.c.}\right]  \nonumber \\
  & +  \frac{1}{4} C_2^{(SD)} \left[ (N^T P_i N)^\dagger (N^T P_j \LRd_x \LRd_y N)(\delta_{ix}\delta_{jy} - \frac{1}{3}\delta_{ij}\delta_{xy}) + {\rm H.c.}\right]  +  \frac{1}{4} C_2^{(\oneP)} (N^T P_0 \LRd_i  N)^\dagger (N^T P_0 \LRd_i  N) \nonumber \\
  & +  \frac{1}{4}  \left[  C_2^{(\Pzero)} \delta_{xy}\delta_{wz}+ C_2^{(\Pone)} (\delta_{xw}\delta_{yz}-\delta_{xz}\delta_{yw}) \right. \nonumber \\
  & \qquad\left. + C_2^{(\Ptwo)} (2\delta_{xw}\delta_{yz}+2\delta_{xz}\delta_{yw}-\frac{4}{3}\delta_{xy}\delta_{wz}) \right]  (N^T P_{y,a} \LRd_x  N)^\dagger (N^T P_{z,a} \LRd_w  N)\,,
\end{align}  
Here $N^T \calO \LRd_i N \equiv N^T \calO \nabla_i N - (\nabla_i N^T) \calO N$, and $\calO$ contains some combination of the identity and Pauli matrices in spin or isospin space.
The projection operators are defined by 
\begin{align} 
  P_i & = \frac{1}{\sqrt{8}} \sigma_2 \sigma_i \tau_2\,, 
  & P_a & = \frac{1}{\sqrt{8}} \sigma_2  \tau_2 \tau_a\,, 
  & P_0 & = \frac{1}{\sqrt{8}} \sigma_2 \tau_2\,, 
  & P_{i,a} & = \frac{1}{\sqrt{8}} \sigma_2 \sigma_i \tau_2 \tau_a\,.
\end{align}
The corresponding LECs are related to those in Eq.~\eqref{eq:two_derivative_ops} according to
\begin{align}\label{NLOrelns}
  C_2^{(\oneS)}  = & -4 (\couplingOneZero - 3 \couplingThreeOne - \couplingSixOne) 
  - 4(\couplingOneOne- 3\couplingThreeZero  - \couplingSixZero ) \\  
  C_2^{(\threeS)}   = & -4 (\couplingOneZero - 3 \couplingThreeOne - \couplingSixOne) -4( -3 \couplingOneOne + \couplingThreeZero  + \frac{1}{3}\couplingSixZero ) \\
  C_2^{(SD)} =& - 4 ( 3\couplingSixOne) -4( - \couplingSixZero) \\
  C_2^{(\oneP)}  = & -4 (\couplingOneZero + 9 \couplingThreeOne + 3 \couplingSixOne) -4( - 3 \couplingOneOne - 3 \couplingThreeZero  - \couplingSixZero ) \\
  C_2^{(\Pzero)}  = & -\frac{4}{3} (\couplingOneZero + \couplingThreeOne - 3 \couplingSixOne) -\frac{4}{3}( \couplingOneOne  + \couplingThreeZero  - 2 \couplingFiveZero  - 3 \couplingSixZero) \\
  C_2^{(\Pone)}  = &-2 (\couplingOneZero + \couplingThreeOne + 2 \couplingSixOne) -2(  \couplingOneOne  + \couplingThreeZero - \couplingFiveZero  + 2 \couplingSixZero) \\
  C_2^{(\Ptwo)}  = &-(\couplingOneZero +  \couplingThreeOne) -(  \couplingOneOne + \couplingThreeZero  + \couplingFiveZero)\,. 
\end{align}            
\end{widetext}
Setting the couplings from Eq.~\eqref{eq:two_derivative_ops} to their Wigner-symmetric values leads to the partial wave couplings in Eq.~\eqref{eq:wigner_couplings}.
Again, this implies that $\calA^{(\oneS)} = \calA^{(\threeS)} \equiv \calA_S$ and $\calA^{(\onePone)} = \calA^{(\PJ)} \equiv \calA_P$ for $J = 0$, $1$, and $2$.
Furthermore, the Wigner limit forces the $S-D$ mixing term to vanish as expected, i.e., $C_2^{(SD)} = 0$.

\bibliographystyle{apsrev4-2}
\bibliography{biblio}

\begin{thebibliography}{75}%
\makeatletter
\providecommand \@ifxundefined [1]{%
 \@ifx{#1\undefined}
}%
\providecommand \@ifnum [1]{%
 \ifnum #1\expandafter \@firstoftwo
 \else \expandafter \@secondoftwo
 \fi
}%
\providecommand \@ifx [1]{%
 \ifx #1\expandafter \@firstoftwo
 \else \expandafter \@secondoftwo
 \fi
}%
\providecommand \natexlab [1]{#1}%
\providecommand \enquote  [1]{``#1''}%
\providecommand \bibnamefont  [1]{#1}%
\providecommand \bibfnamefont [1]{#1}%
\providecommand \citenamefont [1]{#1}%
\providecommand \href@noop [0]{\@secondoftwo}%
\providecommand \href [0]{\begingroup \@sanitize@url \@href}%
\providecommand \@href[1]{\@@startlink{#1}\@@href}%
\providecommand \@@href[1]{\endgroup#1\@@endlink}%
\providecommand \@sanitize@url [0]{\catcode `\\12\catcode `\$12\catcode
  `\&12\catcode `\#12\catcode `\^12\catcode `\_12\catcode `\%12\relax}%
\providecommand \@@startlink[1]{}%
\providecommand \@@endlink[0]{}%
\providecommand \url  [0]{\begingroup\@sanitize@url \@url }%
\providecommand \@url [1]{\endgroup\@href {#1}{\urlprefix }}%
\providecommand \urlprefix  [0]{URL }%
\providecommand \Eprint [0]{\href }%
\providecommand \doibase [0]{https://doi.org/}%
\providecommand \selectlanguage [0]{\@gobble}%
\providecommand \bibinfo  [0]{\@secondoftwo}%
\providecommand \bibfield  [0]{\@secondoftwo}%
\providecommand \translation [1]{[#1]}%
\providecommand \BibitemOpen [0]{}%
\providecommand \bibitemStop [0]{}%
\providecommand \bibitemNoStop [0]{.\EOS\space}%
\providecommand \EOS [0]{\spacefactor3000\relax}%
\providecommand \BibitemShut  [1]{\csname bibitem#1\endcsname}%
\let\auto@bib@innerbib\@empty
\bibitem [{\citenamefont {Wigner}(1937)}]{Wigner:1936dx}%
  \BibitemOpen
  \bibfield  {author} {\bibinfo {author} {\bibfnamefont {E.}~\bibnamefont
  {Wigner}},\ }\href {https://doi.org/10.1103/PhysRev.51.106} {\bibfield
  {journal} {\bibinfo  {journal} {Phys. Rev.}\ }\textbf {\bibinfo {volume}
  {51}},\ \bibinfo {pages} {106} (\bibinfo {year} {1937})}\BibitemShut
  {NoStop}%
\bibitem [{\citenamefont {Kaplan}\ and\ \citenamefont
  {Manohar}(1997)}]{Kaplan:1996rk}%
  \BibitemOpen
  \bibfield  {author} {\bibinfo {author} {\bibfnamefont {D.~B.}\ \bibnamefont
  {Kaplan}}\ and\ \bibinfo {author} {\bibfnamefont {A.~V.}\ \bibnamefont
  {Manohar}},\ }\href {https://doi.org/10.1103/PhysRevC.56.76} {\bibfield
  {journal} {\bibinfo  {journal} {Phys. Rev. C}\ }\textbf {\bibinfo {volume}
  {56}},\ \bibinfo {pages} {76} (\bibinfo {year} {1997})},\ \Eprint
  {https://arxiv.org/abs/nucl-th/9612021} {arXiv:nucl-th/9612021} \BibitemShut
  {NoStop}%
\bibitem [{\citenamefont {Kaplan}\ and\ \citenamefont
  {Savage}(1996)}]{Kaplan:1995yg}%
  \BibitemOpen
  \bibfield  {author} {\bibinfo {author} {\bibfnamefont {D.~B.}\ \bibnamefont
  {Kaplan}}\ and\ \bibinfo {author} {\bibfnamefont {M.~J.}\ \bibnamefont
  {Savage}},\ }\href {https://doi.org/10.1016/0370-2693(95)01277-X} {\bibfield
  {journal} {\bibinfo  {journal} {Phys. Lett. B}\ }\textbf {\bibinfo {volume}
  {365}},\ \bibinfo {pages} {244} (\bibinfo {year} {1996})},\ \Eprint
  {https://arxiv.org/abs/hep-ph/9509371} {arXiv:hep-ph/9509371} \BibitemShut
  {NoStop}%
\bibitem [{\citenamefont {Richardson}\ \emph {et~al.}(2023)\citenamefont
  {Richardson}, \citenamefont {Schindler},\ and\ \citenamefont
  {Springer}}]{Richardson:2022hyj}%
  \BibitemOpen
  \bibfield  {author} {\bibinfo {author} {\bibfnamefont {T.~R.}\ \bibnamefont
  {Richardson}}, \bibinfo {author} {\bibfnamefont {M.~R.}\ \bibnamefont
  {Schindler}},\ and\ \bibinfo {author} {\bibfnamefont {R.~P.}\ \bibnamefont
  {Springer}},\ }\href {https://doi.org/10.1146/annurev-nucl-102020-014052}
  {\bibfield  {journal} {\bibinfo  {journal} {Ann. Rev. Nucl. Part. Sci.}\
  }\textbf {\bibinfo {volume} {73}},\ \bibinfo {pages} {123} (\bibinfo {year}
  {2023})},\ \Eprint {https://arxiv.org/abs/2212.13049} {arXiv:2212.13049
  [nucl-th]} \BibitemShut {NoStop}%
\bibitem [{\citenamefont {Mehen}\ \emph {et~al.}(1999)\citenamefont {Mehen},
  \citenamefont {Stewart},\ and\ \citenamefont {Wise}}]{Mehen:1999qs}%
  \BibitemOpen
  \bibfield  {author} {\bibinfo {author} {\bibfnamefont {T.}~\bibnamefont
  {Mehen}}, \bibinfo {author} {\bibfnamefont {I.~W.}\ \bibnamefont {Stewart}},\
  and\ \bibinfo {author} {\bibfnamefont {M.~B.}\ \bibnamefont {Wise}},\ }\href
  {https://doi.org/10.1103/PhysRevLett.83.931} {\bibfield  {journal} {\bibinfo
  {journal} {Phys. Rev. Lett.}\ }\textbf {\bibinfo {volume} {83}},\ \bibinfo
  {pages} {931} (\bibinfo {year} {1999})},\ \Eprint
  {https://arxiv.org/abs/hep-ph/9902370} {arXiv:hep-ph/9902370} \BibitemShut
  {NoStop}%
\bibitem [{\citenamefont {Beane}\ \emph {et~al.}(2019)\citenamefont {Beane},
  \citenamefont {Kaplan}, \citenamefont {Klco},\ and\ \citenamefont
  {Savage}}]{Beane:2018oxh}%
  \BibitemOpen
  \bibfield  {author} {\bibinfo {author} {\bibfnamefont {S.~R.}\ \bibnamefont
  {Beane}}, \bibinfo {author} {\bibfnamefont {D.~B.}\ \bibnamefont {Kaplan}},
  \bibinfo {author} {\bibfnamefont {N.}~\bibnamefont {Klco}},\ and\ \bibinfo
  {author} {\bibfnamefont {M.~J.}\ \bibnamefont {Savage}},\ }\href
  {https://doi.org/10.1103/PhysRevLett.122.102001} {\bibfield  {journal}
  {\bibinfo  {journal} {Phys. Rev. Lett.}\ }\textbf {\bibinfo {volume} {122}},\
  \bibinfo {pages} {102001} (\bibinfo {year} {2019})},\ \Eprint
  {https://arxiv.org/abs/1812.03138} {arXiv:1812.03138 [nucl-th]} \BibitemShut
  {NoStop}%
\bibitem [{\citenamefont {Kaplan}\ \emph {et~al.}(1996)\citenamefont {Kaplan},
  \citenamefont {Savage},\ and\ \citenamefont {Wise}}]{Kaplan:1996xu}%
  \BibitemOpen
  \bibfield  {author} {\bibinfo {author} {\bibfnamefont {D.~B.}\ \bibnamefont
  {Kaplan}}, \bibinfo {author} {\bibfnamefont {M.~J.}\ \bibnamefont {Savage}},\
  and\ \bibinfo {author} {\bibfnamefont {M.~B.}\ \bibnamefont {Wise}},\ }\href
  {https://doi.org/10.1016/0550-3213(96)00357-4} {\bibfield  {journal}
  {\bibinfo  {journal} {Nucl. Phys. B}\ }\textbf {\bibinfo {volume} {478}},\
  \bibinfo {pages} {629} (\bibinfo {year} {1996})},\ \Eprint
  {https://arxiv.org/abs/nucl-th/9605002} {arXiv:nucl-th/9605002} \BibitemShut
  {NoStop}%
\bibitem [{\citenamefont {Kaplan}\ \emph
  {et~al.}(1998{\natexlab{a}})\citenamefont {Kaplan}, \citenamefont {Savage},\
  and\ \citenamefont {Wise}}]{Kaplan:1998tg}%
  \BibitemOpen
  \bibfield  {author} {\bibinfo {author} {\bibfnamefont {D.~B.}\ \bibnamefont
  {Kaplan}}, \bibinfo {author} {\bibfnamefont {M.~J.}\ \bibnamefont {Savage}},\
  and\ \bibinfo {author} {\bibfnamefont {M.~B.}\ \bibnamefont {Wise}},\ }\href
  {https://doi.org/10.1016/S0370-2693(98)00210-X} {\bibfield  {journal}
  {\bibinfo  {journal} {Phys. Lett. B}\ }\textbf {\bibinfo {volume} {424}},\
  \bibinfo {pages} {390} (\bibinfo {year} {1998}{\natexlab{a}})},\ \Eprint
  {https://arxiv.org/abs/nucl-th/9801034} {arXiv:nucl-th/9801034} \BibitemShut
  {NoStop}%
\bibitem [{\citenamefont {Kaplan}\ \emph
  {et~al.}(1998{\natexlab{b}})\citenamefont {Kaplan}, \citenamefont {Savage},\
  and\ \citenamefont {Wise}}]{Kaplan:1998we}%
  \BibitemOpen
  \bibfield  {author} {\bibinfo {author} {\bibfnamefont {D.~B.}\ \bibnamefont
  {Kaplan}}, \bibinfo {author} {\bibfnamefont {M.~J.}\ \bibnamefont {Savage}},\
  and\ \bibinfo {author} {\bibfnamefont {M.~B.}\ \bibnamefont {Wise}},\ }\href
  {https://doi.org/10.1016/S0550-3213(98)00440-4} {\bibfield  {journal}
  {\bibinfo  {journal} {Nucl. Phys. B}\ }\textbf {\bibinfo {volume} {534}},\
  \bibinfo {pages} {329} (\bibinfo {year} {1998}{\natexlab{b}})},\ \Eprint
  {https://arxiv.org/abs/nucl-th/9802075} {arXiv:nucl-th/9802075} \BibitemShut
  {NoStop}%
\bibitem [{\citenamefont {van Kolck}(1999)}]{vanKolck:1998bw}%
  \BibitemOpen
  \bibfield  {author} {\bibinfo {author} {\bibfnamefont {U.}~\bibnamefont {van
  Kolck}},\ }\href {https://doi.org/10.1016/S0375-9474(98)00612-5} {\bibfield
  {journal} {\bibinfo  {journal} {Nucl. Phys. A}\ }\textbf {\bibinfo {volume}
  {645}},\ \bibinfo {pages} {273} (\bibinfo {year} {1999})},\ \Eprint
  {https://arxiv.org/abs/nucl-th/9808007} {arXiv:nucl-th/9808007} \BibitemShut
  {NoStop}%
\bibitem [{\citenamefont {Chen}\ \emph {et~al.}(1999)\citenamefont {Chen},
  \citenamefont {Rupak},\ and\ \citenamefont {Savage}}]{Chen:1999tn}%
  \BibitemOpen
  \bibfield  {author} {\bibinfo {author} {\bibfnamefont {J.-W.}\ \bibnamefont
  {Chen}}, \bibinfo {author} {\bibfnamefont {G.}~\bibnamefont {Rupak}},\ and\
  \bibinfo {author} {\bibfnamefont {M.~J.}\ \bibnamefont {Savage}},\ }\href
  {https://doi.org/10.1016/S0375-9474(99)00298-5} {\bibfield  {journal}
  {\bibinfo  {journal} {Nucl. Phys. A}\ }\textbf {\bibinfo {volume} {653}},\
  \bibinfo {pages} {386} (\bibinfo {year} {1999})},\ \Eprint
  {https://arxiv.org/abs/nucl-th/9902056} {arXiv:nucl-th/9902056} \BibitemShut
  {NoStop}%
\bibitem [{\citenamefont {Mehen}\ \emph {et~al.}(2000)\citenamefont {Mehen},
  \citenamefont {Stewart},\ and\ \citenamefont {Wise}}]{Mehen:1999nd}%
  \BibitemOpen
  \bibfield  {author} {\bibinfo {author} {\bibfnamefont {T.}~\bibnamefont
  {Mehen}}, \bibinfo {author} {\bibfnamefont {I.~W.}\ \bibnamefont {Stewart}},\
  and\ \bibinfo {author} {\bibfnamefont {M.~B.}\ \bibnamefont {Wise}},\ }\href
  {https://doi.org/10.1016/S0370-2693(00)00006-X} {\bibfield  {journal}
  {\bibinfo  {journal} {Phys. Lett. B}\ }\textbf {\bibinfo {volume} {474}},\
  \bibinfo {pages} {145} (\bibinfo {year} {2000})},\ \Eprint
  {https://arxiv.org/abs/hep-th/9910025} {arXiv:hep-th/9910025} \BibitemShut
  {NoStop}%
\bibitem [{\citenamefont {Low}\ and\ \citenamefont
  {Mehen}(2021)}]{Low:2021ufv}%
  \BibitemOpen
  \bibfield  {author} {\bibinfo {author} {\bibfnamefont {I.}~\bibnamefont
  {Low}}\ and\ \bibinfo {author} {\bibfnamefont {T.}~\bibnamefont {Mehen}},\
  }\href {https://doi.org/10.1103/PhysRevD.104.074014} {\bibfield  {journal}
  {\bibinfo  {journal} {Phys. Rev. D}\ }\textbf {\bibinfo {volume} {104}},\
  \bibinfo {pages} {074014} (\bibinfo {year} {2021})},\ \Eprint
  {https://arxiv.org/abs/2104.10835} {arXiv:2104.10835 [hep-th]} \BibitemShut
  {NoStop}%
\bibitem [{\citenamefont {Liu}\ \emph {et~al.}(2023)\citenamefont {Liu},
  \citenamefont {Low},\ and\ \citenamefont {Mehen}}]{Liu:2022grf}%
  \BibitemOpen
  \bibfield  {author} {\bibinfo {author} {\bibfnamefont {Q.}~\bibnamefont
  {Liu}}, \bibinfo {author} {\bibfnamefont {I.}~\bibnamefont {Low}},\ and\
  \bibinfo {author} {\bibfnamefont {T.}~\bibnamefont {Mehen}},\ }\href
  {https://doi.org/10.1103/PhysRevC.107.025204} {\bibfield  {journal} {\bibinfo
   {journal} {Phys. Rev. C}\ }\textbf {\bibinfo {volume} {107}},\ \bibinfo
  {pages} {025204} (\bibinfo {year} {2023})},\ \Eprint
  {https://arxiv.org/abs/2210.12085} {arXiv:2210.12085 [quant-ph]} \BibitemShut
  {NoStop}%
\bibitem [{\citenamefont {Beane}\ \emph {et~al.}(2021)\citenamefont {Beane},
  \citenamefont {Farrell},\ and\ \citenamefont {Varma}}]{Beane:2021zvo}%
  \BibitemOpen
  \bibfield  {author} {\bibinfo {author} {\bibfnamefont {S.~R.}\ \bibnamefont
  {Beane}}, \bibinfo {author} {\bibfnamefont {R.~C.}\ \bibnamefont {Farrell}},\
  and\ \bibinfo {author} {\bibfnamefont {M.}~\bibnamefont {Varma}},\ }\href
  {https://doi.org/10.1142/S0217751X21502055} {\bibfield  {journal} {\bibinfo
  {journal} {Int. J. Mod. Phys. A}\ }\textbf {\bibinfo {volume} {36}},\
  \bibinfo {pages} {2150205} (\bibinfo {year} {2021})},\ \Eprint
  {https://arxiv.org/abs/2108.00646} {arXiv:2108.00646 [hep-ph]} \BibitemShut
  {NoStop}%
\bibitem [{\citenamefont {Hu}\ \emph {et~al.}(2024)\citenamefont {Hu},
  \citenamefont {Chen},\ and\ \citenamefont {Guo}}]{Hu:2024hex}%
  \BibitemOpen
  \bibfield  {author} {\bibinfo {author} {\bibfnamefont {T.-R.}\ \bibnamefont
  {Hu}}, \bibinfo {author} {\bibfnamefont {S.}~\bibnamefont {Chen}},\ and\
  \bibinfo {author} {\bibfnamefont {F.-K.}\ \bibnamefont {Guo}},\ }\href
  {https://doi.org/10.1103/PhysRevD.110.014001} {\bibfield  {journal} {\bibinfo
   {journal} {Phys. Rev. D}\ }\textbf {\bibinfo {volume} {110}},\ \bibinfo
  {pages} {014001} (\bibinfo {year} {2024})},\ \Eprint
  {https://arxiv.org/abs/2404.05958} {arXiv:2404.05958 [hep-ph]} \BibitemShut
  {NoStop}%
\bibitem [{\citenamefont {Hu}\ \emph {et~al.}(2025)\citenamefont {Hu},
  \citenamefont {Sone}, \citenamefont {Guo}, \citenamefont {Hyodo},\ and\
  \citenamefont {Low}}]{Hu:2025lua}%
  \BibitemOpen
  \bibfield  {author} {\bibinfo {author} {\bibfnamefont {T.-R.}\ \bibnamefont
  {Hu}}, \bibinfo {author} {\bibfnamefont {K.}~\bibnamefont {Sone}}, \bibinfo
  {author} {\bibfnamefont {F.-K.}\ \bibnamefont {Guo}}, \bibinfo {author}
  {\bibfnamefont {T.}~\bibnamefont {Hyodo}},\ and\ \bibinfo {author}
  {\bibfnamefont {I.}~\bibnamefont {Low}},\ }\href
  {https://doi.org/10.1103/x52w-4rbs} {\bibfield  {journal} {\bibinfo
  {journal} {Phys. Rev. Res.}\ }\textbf {\bibinfo {volume} {7}},\ \bibinfo
  {pages} {043306} (\bibinfo {year} {2025})},\ \Eprint
  {https://arxiv.org/abs/2506.08960} {arXiv:2506.08960 [hep-ph]} \BibitemShut
  {NoStop}%
\bibitem [{\citenamefont {Hu}\ \emph {et~al.}(2026)\citenamefont {Hu},
  \citenamefont {Chen}, \citenamefont {Sone}, \citenamefont {Guo},
  \citenamefont {Hyodo},\ and\ \citenamefont {Low}}]{Hu:2025jne}%
  \BibitemOpen
  \bibfield  {author} {\bibinfo {author} {\bibfnamefont {T.-R.}\ \bibnamefont
  {Hu}}, \bibinfo {author} {\bibfnamefont {S.}~\bibnamefont {Chen}}, \bibinfo
  {author} {\bibfnamefont {K.}~\bibnamefont {Sone}}, \bibinfo {author}
  {\bibfnamefont {F.-K.}\ \bibnamefont {Guo}}, \bibinfo {author} {\bibfnamefont
  {T.}~\bibnamefont {Hyodo}},\ and\ \bibinfo {author} {\bibfnamefont
  {I.}~\bibnamefont {Low}},\ }\href {https://doi.org/10.22323/1.500.0104}
  {\bibfield  {journal} {\bibinfo  {journal} {PoS}\ }\textbf {\bibinfo {volume}
  {HADRON2025}},\ \bibinfo {pages} {104} (\bibinfo {year} {2026})},\ \Eprint
  {https://arxiv.org/abs/2507.22694} {arXiv:2507.22694 [hep-ph]} \BibitemShut
  {NoStop}%
\bibitem [{\citenamefont {Sone}\ \emph {et~al.}(2026)\citenamefont {Sone},
  \citenamefont {Hu}, \citenamefont {Guo}, \citenamefont {Hyodo},\ and\
  \citenamefont {Low}}]{Sone:2026jmo}%
  \BibitemOpen
  \bibfield  {author} {\bibinfo {author} {\bibfnamefont {K.}~\bibnamefont
  {Sone}}, \bibinfo {author} {\bibfnamefont {T.-R.}\ \bibnamefont {Hu}},
  \bibinfo {author} {\bibfnamefont {F.-K.}\ \bibnamefont {Guo}}, \bibinfo
  {author} {\bibfnamefont {T.}~\bibnamefont {Hyodo}},\ and\ \bibinfo {author}
  {\bibfnamefont {I.}~\bibnamefont {Low}}\ }(\bibinfo {year} {2026})\ \Eprint
  {https://arxiv.org/abs/2602.09630} {arXiv:2602.09630 [hep-ph]} \BibitemShut
  {NoStop}%
\bibitem [{\citenamefont {Low}\ and\ \citenamefont {Yin}(2025)}]{Low:2024hvn}%
  \BibitemOpen
  \bibfield  {author} {\bibinfo {author} {\bibfnamefont {I.}~\bibnamefont
  {Low}}\ and\ \bibinfo {author} {\bibfnamefont {Z.}~\bibnamefont {Yin}},\
  }\href {https://doi.org/10.1103/PhysRevD.111.065027} {\bibfield  {journal}
  {\bibinfo  {journal} {Phys. Rev. D}\ }\textbf {\bibinfo {volume} {111}},\
  \bibinfo {pages} {065027} (\bibinfo {year} {2025})},\ \Eprint
  {https://arxiv.org/abs/2410.22414} {arXiv:2410.22414 [hep-th]} \BibitemShut
  {NoStop}%
\bibitem [{\citenamefont {Low}\ and\ \citenamefont {Yin}(2026)}]{Low:2024mrk}%
  \BibitemOpen
  \bibfield  {author} {\bibinfo {author} {\bibfnamefont {I.}~\bibnamefont
  {Low}}\ and\ \bibinfo {author} {\bibfnamefont {Z.}~\bibnamefont {Yin}},\
  }\href {https://doi.org/10.1103/3yg7-r5s9} {\bibfield  {journal} {\bibinfo
  {journal} {Phys. Rev. D}\ }\textbf {\bibinfo {volume} {113}},\ \bibinfo
  {pages} {065004} (\bibinfo {year} {2026})},\ \Eprint
  {https://arxiv.org/abs/2405.08056} {arXiv:2405.08056 [hep-th]} \BibitemShut
  {NoStop}%
\bibitem [{\citenamefont {McGinnis}(2025)}]{McGinnis:2025brt}%
  \BibitemOpen
  \bibfield  {author} {\bibinfo {author} {\bibfnamefont {N.}~\bibnamefont
  {McGinnis}},\ }\href@noop {} {\bibinfo {title} {{Symmetry, entanglement, and
  the $S$-matrix}}} (\bibinfo {year} {2025}),\ \Eprint
  {https://arxiv.org/abs/2504.21079} {arXiv:2504.21079 [hep-th]} \BibitemShut
  {NoStop}%
\bibitem [{\citenamefont {Cervera-Lierta}\ \emph {et~al.}(2017)\citenamefont
  {Cervera-Lierta}, \citenamefont {Latorre}, \citenamefont {Rojo},\ and\
  \citenamefont {Rottoli}}]{Cervera-Lierta:2017tdt}%
  \BibitemOpen
  \bibfield  {author} {\bibinfo {author} {\bibfnamefont {A.}~\bibnamefont
  {Cervera-Lierta}}, \bibinfo {author} {\bibfnamefont {J.~I.}\ \bibnamefont
  {Latorre}}, \bibinfo {author} {\bibfnamefont {J.}~\bibnamefont {Rojo}},\ and\
  \bibinfo {author} {\bibfnamefont {L.}~\bibnamefont {Rottoli}},\ }\href
  {https://doi.org/10.21468/SciPostPhys.3.5.036} {\bibfield  {journal}
  {\bibinfo  {journal} {SciPost Phys.}\ }\textbf {\bibinfo {volume} {3}},\
  \bibinfo {pages} {036} (\bibinfo {year} {2017})},\ \Eprint
  {https://arxiv.org/abs/1703.02989} {arXiv:1703.02989 [hep-th]} \BibitemShut
  {NoStop}%
\bibitem [{\citenamefont {Carena}\ \emph {et~al.}(2024)\citenamefont {Carena},
  \citenamefont {Low}, \citenamefont {Wagner},\ and\ \citenamefont
  {Xiao}}]{Carena:2023vjc}%
  \BibitemOpen
  \bibfield  {author} {\bibinfo {author} {\bibfnamefont {M.}~\bibnamefont
  {Carena}}, \bibinfo {author} {\bibfnamefont {I.}~\bibnamefont {Low}},
  \bibinfo {author} {\bibfnamefont {C.~E.~M.}\ \bibnamefont {Wagner}},\ and\
  \bibinfo {author} {\bibfnamefont {M.-L.}\ \bibnamefont {Xiao}},\ }\href
  {https://doi.org/10.1103/PhysRevD.109.L051901} {\bibfield  {journal}
  {\bibinfo  {journal} {Phys. Rev. D}\ }\textbf {\bibinfo {volume} {109}},\
  \bibinfo {pages} {L051901} (\bibinfo {year} {2024})},\ \Eprint
  {https://arxiv.org/abs/2307.08112} {arXiv:2307.08112 [hep-ph]} \BibitemShut
  {NoStop}%
\bibitem [{\citenamefont {Chang}\ and\ \citenamefont
  {Jacobo}(2024)}]{Chang:2024wrx}%
  \BibitemOpen
  \bibfield  {author} {\bibinfo {author} {\bibfnamefont {S.}~\bibnamefont
  {Chang}}\ and\ \bibinfo {author} {\bibfnamefont {G.}~\bibnamefont {Jacobo}},\
  }\href {https://doi.org/10.1103/PhysRevD.110.096020} {\bibfield  {journal}
  {\bibinfo  {journal} {Phys. Rev. D}\ }\textbf {\bibinfo {volume} {110}},\
  \bibinfo {pages} {096020} (\bibinfo {year} {2024})},\ \Eprint
  {https://arxiv.org/abs/2409.13030} {arXiv:2409.13030 [hep-ph]} \BibitemShut
  {NoStop}%
\bibitem [{\citenamefont {Kowalska}\ and\ \citenamefont
  {Sessolo}(2024)}]{Kowalska:2024kbs}%
  \BibitemOpen
  \bibfield  {author} {\bibinfo {author} {\bibfnamefont {K.}~\bibnamefont
  {Kowalska}}\ and\ \bibinfo {author} {\bibfnamefont {E.~M.}\ \bibnamefont
  {Sessolo}},\ }\href {https://doi.org/10.1007/JHEP07(2024)156} {\bibfield
  {journal} {\bibinfo  {journal} {JHEP}\ }\textbf {\bibinfo {volume} {07}},\
  \bibinfo {pages} {156}},\ \Eprint {https://arxiv.org/abs/2404.13743}
  {arXiv:2404.13743 [hep-ph]} \BibitemShut {NoStop}%
\bibitem [{\citenamefont {Carena}\ \emph {et~al.}(2025)\citenamefont {Carena},
  \citenamefont {Coloretti}, \citenamefont {Liu}, \citenamefont {Littmann},
  \citenamefont {Low},\ and\ \citenamefont {Wagner}}]{Carena:2025wyh}%
  \BibitemOpen
  \bibfield  {author} {\bibinfo {author} {\bibfnamefont {M.}~\bibnamefont
  {Carena}}, \bibinfo {author} {\bibfnamefont {G.}~\bibnamefont {Coloretti}},
  \bibinfo {author} {\bibfnamefont {W.}~\bibnamefont {Liu}}, \bibinfo {author}
  {\bibfnamefont {M.}~\bibnamefont {Littmann}}, \bibinfo {author}
  {\bibfnamefont {I.}~\bibnamefont {Low}},\ and\ \bibinfo {author}
  {\bibfnamefont {C.~E.~M.}\ \bibnamefont {Wagner}},\ }\href
  {https://doi.org/10.1007/JHEP08(2025)016} {\bibfield  {journal} {\bibinfo
  {journal} {JHEP}\ }\textbf {\bibinfo {volume} {08}},\ \bibinfo {pages}
  {016}},\ \Eprint {https://arxiv.org/abs/2505.00873} {arXiv:2505.00873
  [hep-ph]} \BibitemShut {NoStop}%
\bibitem [{\citenamefont {Liu}\ \emph {et~al.}(2025{\natexlab{a}})\citenamefont
  {Liu}, \citenamefont {Tanaka}, \citenamefont {Wang}, \citenamefont {Zhang},\
  and\ \citenamefont {Zheng}}]{Liu:2025iwh}%
  \BibitemOpen
  \bibfield  {author} {\bibinfo {author} {\bibfnamefont {J.}~\bibnamefont
  {Liu}}, \bibinfo {author} {\bibfnamefont {M.}~\bibnamefont {Tanaka}},
  \bibinfo {author} {\bibfnamefont {X.-P.}\ \bibnamefont {Wang}}, \bibinfo
  {author} {\bibfnamefont {J.-J.}\ \bibnamefont {Zhang}},\ and\ \bibinfo
  {author} {\bibfnamefont {Z.}~\bibnamefont {Zheng}},\ }\href@noop {} {\bibinfo
  {title} {{Parameter Inference from Final-State Entanglement in Higgs
  Decays}}} (\bibinfo {year} {2025}{\natexlab{a}}),\ \Eprint
  {https://arxiv.org/abs/2511.17321} {arXiv:2511.17321 [hep-ph]} \BibitemShut
  {NoStop}%
\bibitem [{\citenamefont {Thaler}\ and\ \citenamefont
  {Trifinopoulos}(2025)}]{Thaler:2024anb}%
  \BibitemOpen
  \bibfield  {author} {\bibinfo {author} {\bibfnamefont {J.}~\bibnamefont
  {Thaler}}\ and\ \bibinfo {author} {\bibfnamefont {S.}~\bibnamefont
  {Trifinopoulos}},\ }\href {https://doi.org/10.1103/PhysRevD.111.056021}
  {\bibfield  {journal} {\bibinfo  {journal} {Phys. Rev. D}\ }\textbf {\bibinfo
  {volume} {111}},\ \bibinfo {pages} {056021} (\bibinfo {year} {2025})},\
  \Eprint {https://arxiv.org/abs/2410.23343} {arXiv:2410.23343 [hep-ph]}
  \BibitemShut {NoStop}%
\bibitem [{\citenamefont {Liu}\ \emph {et~al.}(2025{\natexlab{b}})\citenamefont
  {Liu}, \citenamefont {Low},\ and\ \citenamefont {Yin}}]{Liu:2025bgw}%
  \BibitemOpen
  \bibfield  {author} {\bibinfo {author} {\bibfnamefont {Q.}~\bibnamefont
  {Liu}}, \bibinfo {author} {\bibfnamefont {I.}~\bibnamefont {Low}},\ and\
  \bibinfo {author} {\bibfnamefont {Z.}~\bibnamefont {Yin}},\ }\href@noop {}
  {\bibinfo {title} {{A Quantum Computational Determination of the Weak Mixing
  Angle in the Standard Model}}} (\bibinfo {year} {2025}{\natexlab{b}}),\
  \Eprint {https://arxiv.org/abs/2509.18251} {arXiv:2509.18251 [hep-ph]}
  \BibitemShut {NoStop}%
\bibitem [{\citenamefont {Li}\ \emph {et~al.}(2026)\citenamefont {Li},
  \citenamefont {Ma}, \citenamefont {Shu},\ and\ \citenamefont
  {Zhu}}]{Li:2026kha}%
  \BibitemOpen
  \bibfield  {author} {\bibinfo {author} {\bibfnamefont {C.}~\bibnamefont
  {Li}}, \bibinfo {author} {\bibfnamefont {T.}~\bibnamefont {Ma}}, \bibinfo
  {author} {\bibfnamefont {J.}~\bibnamefont {Shu}},\ and\ \bibinfo {author}
  {\bibfnamefont {M.}~\bibnamefont {Zhu}},\ }\href@noop {} {\bibinfo {title}
  {{Entanglement Maximization and Symmetry Selection in Composite Higgs
  Models}}} (\bibinfo {year} {2026}),\ \Eprint
  {https://arxiv.org/abs/2605.17434} {arXiv:2605.17434 [hep-ph]} \BibitemShut
  {NoStop}%
\bibitem [{\citenamefont {Cao}\ \emph {et~al.}(2026)\citenamefont {Cao},
  \citenamefont {Liu}, \citenamefont {Qi}, \citenamefont {Zhang},\ and\
  \citenamefont {Zhao}}]{Cao:2026aye}%
  \BibitemOpen
  \bibfield  {author} {\bibinfo {author} {\bibfnamefont {Q.-H.}\ \bibnamefont
  {Cao}}, \bibinfo {author} {\bibfnamefont {Y.}~\bibnamefont {Liu}}, \bibinfo
  {author} {\bibfnamefont {H.}~\bibnamefont {Qi}}, \bibinfo {author}
  {\bibfnamefont {H.}~\bibnamefont {Zhang}},\ and\ \bibinfo {author}
  {\bibfnamefont {H.}~\bibnamefont {Zhao}},\ }\href@noop {} {\bibinfo {title}
  {{Symmetry Breaking as Quantum Gate: Entropy and Weak Mixing Angle}}}
  (\bibinfo {year} {2026}),\ \Eprint {https://arxiv.org/abs/2605.22070}
  {arXiv:2605.22070 [hep-ph]} \BibitemShut {NoStop}%
\bibitem [{\citenamefont {Low}\ and\ \citenamefont
  {Goswami}(2026)}]{Low:2026oyf}%
  \BibitemOpen
  \bibfield  {author} {\bibinfo {author} {\bibfnamefont {I.}~\bibnamefont
  {Low}}\ and\ \bibinfo {author} {\bibfnamefont {P.}~\bibnamefont {Goswami}},\
  }\href@noop {} {\bibinfo {title} {{Entangling Power: A Probe of Symmetry and
  Integrability in Quantum Many-Body Systems}}} (\bibinfo {year} {2026}),\
  \Eprint {https://arxiv.org/abs/2605.20661} {arXiv:2605.20661 [quant-ph]}
  \BibitemShut {NoStop}%
\bibitem [{\citenamefont {Bravyi}\ and\ \citenamefont
  {Kitaev}(2005)}]{Bravyi:2004isx}%
  \BibitemOpen
  \bibfield  {author} {\bibinfo {author} {\bibfnamefont {S.}~\bibnamefont
  {Bravyi}}\ and\ \bibinfo {author} {\bibfnamefont {A.}~\bibnamefont
  {Kitaev}},\ }\href {https://doi.org/10.1103/PhysRevA.71.022316} {\bibfield
  {journal} {\bibinfo  {journal} {Phys. Rev. A}\ }\textbf {\bibinfo {volume}
  {71}},\ \bibinfo {pages} {022316} (\bibinfo {year} {2005})},\ \Eprint
  {https://arxiv.org/abs/quant-ph/0403025} {arXiv:quant-ph/0403025}
  \BibitemShut {NoStop}%
\bibitem [{\citenamefont {Emerson}\ \emph {et~al.}(2014)\citenamefont
  {Emerson}, \citenamefont {Gottesman}, \citenamefont {Mousavian},\ and\
  \citenamefont {Veitch}}]{Emerson:2013zse}%
  \BibitemOpen
  \bibfield  {author} {\bibinfo {author} {\bibfnamefont {J.}~\bibnamefont
  {Emerson}}, \bibinfo {author} {\bibfnamefont {D.}~\bibnamefont {Gottesman}},
  \bibinfo {author} {\bibfnamefont {S.~A.~H.}\ \bibnamefont {Mousavian}},\ and\
  \bibinfo {author} {\bibfnamefont {V.}~\bibnamefont {Veitch}},\ }\href
  {https://doi.org/10.1088/1367-2630/16/1/013009} {\bibfield  {journal}
  {\bibinfo  {journal} {New J. Phys.}\ }\textbf {\bibinfo {volume} {16}},\
  \bibinfo {pages} {013009} (\bibinfo {year} {2014})},\ \Eprint
  {https://arxiv.org/abs/1307.7171} {arXiv:1307.7171 [quant-ph]} \BibitemShut
  {NoStop}%
\bibitem [{\citenamefont {Qian}\ and\ \citenamefont
  {Wang}(2025)}]{Qian:2025oit}%
  \BibitemOpen
  \bibfield  {author} {\bibinfo {author} {\bibfnamefont {D.}~\bibnamefont
  {Qian}}\ and\ \bibinfo {author} {\bibfnamefont {J.}~\bibnamefont {Wang}},\
  }\href {https://doi.org/10.1103/PhysRevA.111.052443} {\bibfield  {journal}
  {\bibinfo  {journal} {Phys. Rev. A}\ }\textbf {\bibinfo {volume} {111}},\
  \bibinfo {pages} {052443} (\bibinfo {year} {2025})},\ \Eprint
  {https://arxiv.org/abs/2502.06393} {arXiv:2502.06393 [quant-ph]} \BibitemShut
  {NoStop}%
\bibitem [{\citenamefont {Gargalionis}\ \emph
  {et~al.}(2026{\natexlab{a}})\citenamefont {Gargalionis}, \citenamefont
  {Moynihan}, \citenamefont {Reichenberg~Ashby}, \citenamefont {Wallace},
  \citenamefont {White},\ and\ \citenamefont {White}}]{Gargalionis:2026onv}%
  \BibitemOpen
  \bibfield  {author} {\bibinfo {author} {\bibfnamefont {J.}~\bibnamefont
  {Gargalionis}}, \bibinfo {author} {\bibfnamefont {N.}~\bibnamefont
  {Moynihan}}, \bibinfo {author} {\bibfnamefont {M.~L.}\ \bibnamefont
  {Reichenberg~Ashby}}, \bibinfo {author} {\bibfnamefont {E.~N.~V.}\
  \bibnamefont {Wallace}}, \bibinfo {author} {\bibfnamefont {C.~D.}\
  \bibnamefont {White}},\ and\ \bibinfo {author} {\bibfnamefont {M.~J.}\
  \bibnamefont {White}},\ }\href@noop {} {\bibinfo {title} {{Non-local
  nonstabiliserness in Gluon and Graviton Scattering}}} (\bibinfo {year}
  {2026}{\natexlab{a}}),\ \Eprint {https://arxiv.org/abs/2603.04148}
  {arXiv:2603.04148 [hep-th]} \BibitemShut {NoStop}%
\bibitem [{\citenamefont {Busoni}\ \emph {et~al.}(2026)\citenamefont {Busoni},
  \citenamefont {Gargalionis}, \citenamefont {Wallace},\ and\ \citenamefont
  {White}}]{Busoni:2026lvp}%
  \BibitemOpen
  \bibfield  {author} {\bibinfo {author} {\bibfnamefont {G.}~\bibnamefont
  {Busoni}}, \bibinfo {author} {\bibfnamefont {J.}~\bibnamefont {Gargalionis}},
  \bibinfo {author} {\bibfnamefont {E.~N.~V.}\ \bibnamefont {Wallace}},\ and\
  \bibinfo {author} {\bibfnamefont {M.~J.}\ \bibnamefont {White}},\ }\href@noop
  {} {\bibinfo {title} {{Analytic formulae for non-local magic in bipartite
  systems of qutrits and ququints}}} (\bibinfo {year} {2026}),\ \Eprint
  {https://arxiv.org/abs/2603.09155} {arXiv:2603.09155 [quant-ph]} \BibitemShut
  {NoStop}%
\bibitem [{\citenamefont {Chernyshev}\ \emph {et~al.}(2025)\citenamefont
  {Chernyshev}, \citenamefont {Robin},\ and\ \citenamefont
  {Savage}}]{Chernyshev:2024pqy}%
  \BibitemOpen
  \bibfield  {author} {\bibinfo {author} {\bibfnamefont {I.}~\bibnamefont
  {Chernyshev}}, \bibinfo {author} {\bibfnamefont {C.~E.~P.}\ \bibnamefont
  {Robin}},\ and\ \bibinfo {author} {\bibfnamefont {M.~J.}\ \bibnamefont
  {Savage}},\ }\href {https://doi.org/10.1103/PhysRevResearch.7.023228}
  {\bibfield  {journal} {\bibinfo  {journal} {Phys. Rev. Res.}\ }\textbf
  {\bibinfo {volume} {7}},\ \bibinfo {pages} {023228} (\bibinfo {year}
  {2025})},\ \Eprint {https://arxiv.org/abs/2411.04203} {arXiv:2411.04203
  [quant-ph]} \BibitemShut {NoStop}%
\bibitem [{\citenamefont {White}\ and\ \citenamefont
  {White}(2024)}]{White:2024nuc}%
  \BibitemOpen
  \bibfield  {author} {\bibinfo {author} {\bibfnamefont {C.~D.}\ \bibnamefont
  {White}}\ and\ \bibinfo {author} {\bibfnamefont {M.~J.}\ \bibnamefont
  {White}},\ }\href {https://doi.org/10.1103/PhysRevD.110.116016} {\bibfield
  {journal} {\bibinfo  {journal} {Phys. Rev. D}\ }\textbf {\bibinfo {volume}
  {110}},\ \bibinfo {pages} {116016} (\bibinfo {year} {2024})},\ \Eprint
  {https://arxiv.org/abs/2406.07321} {arXiv:2406.07321 [hep-ph]} \BibitemShut
  {NoStop}%
\bibitem [{\citenamefont {Aoude}\ \emph {et~al.}(2025)\citenamefont {Aoude},
  \citenamefont {Banks}, \citenamefont {White},\ and\ \citenamefont
  {White}}]{Aoude:2025jzc}%
  \BibitemOpen
  \bibfield  {author} {\bibinfo {author} {\bibfnamefont {R.}~\bibnamefont
  {Aoude}}, \bibinfo {author} {\bibfnamefont {H.}~\bibnamefont {Banks}},
  \bibinfo {author} {\bibfnamefont {C.~D.}\ \bibnamefont {White}},\ and\
  \bibinfo {author} {\bibfnamefont {M.~J.}\ \bibnamefont {White}},\ }\href@noop
  {} {\bibinfo {title} {{Probing new physics in the top sector using quantum
  information}}} (\bibinfo {year} {2025}),\ \Eprint
  {https://arxiv.org/abs/2505.12522} {arXiv:2505.12522 [hep-ph]} \BibitemShut
  {NoStop}%
\bibitem [{\citenamefont {Liu}\ \emph {et~al.}(2026{\natexlab{a}})\citenamefont
  {Liu}, \citenamefont {Low},\ and\ \citenamefont {Yin}}]{Liu:2025qfl}%
  \BibitemOpen
  \bibfield  {author} {\bibinfo {author} {\bibfnamefont {Q.}~\bibnamefont
  {Liu}}, \bibinfo {author} {\bibfnamefont {I.}~\bibnamefont {Low}},\ and\
  \bibinfo {author} {\bibfnamefont {Z.}~\bibnamefont {Yin}},\ }\href
  {https://doi.org/10.1103/l8vw-kwhz} {\bibfield  {journal} {\bibinfo
  {journal} {Phys. Rev. D}\ }\textbf {\bibinfo {volume} {113}},\ \bibinfo
  {pages} {056001} (\bibinfo {year} {2026}{\natexlab{a}})},\ \Eprint
  {https://arxiv.org/abs/2503.03098} {arXiv:2503.03098 [quant-ph]} \BibitemShut
  {NoStop}%
\bibitem [{\citenamefont {Chang}\ \emph {et~al.}(2026)\citenamefont {Chang},
  \citenamefont {Driscoll},\ and\ \citenamefont
  {Trifinopoulos}}]{Chang:2026nlo}%
  \BibitemOpen
  \bibfield  {author} {\bibinfo {author} {\bibfnamefont {S.}~\bibnamefont
  {Chang}}, \bibinfo {author} {\bibfnamefont {T.}~\bibnamefont {Driscoll}},\
  and\ \bibinfo {author} {\bibfnamefont {S.}~\bibnamefont {Trifinopoulos}},\
  }\href@noop {} {\bibinfo {title} {{Next-to-Leading-Order Electroweak
  Corrections to Quantum Resources in Lepton-Lepton Colliders}}} (\bibinfo
  {year} {2026}),\ \bibinfo {note} {in preparation}\BibitemShut {NoStop}%
\bibitem [{\citenamefont {Busoni}\ \emph {et~al.}(2025)\citenamefont {Busoni},
  \citenamefont {Gargalionis}, \citenamefont {Wallace},\ and\ \citenamefont
  {White}}]{Busoni:2025dns}%
  \BibitemOpen
  \bibfield  {author} {\bibinfo {author} {\bibfnamefont {G.}~\bibnamefont
  {Busoni}}, \bibinfo {author} {\bibfnamefont {J.}~\bibnamefont {Gargalionis}},
  \bibinfo {author} {\bibfnamefont {E.~N.~V.}\ \bibnamefont {Wallace}},\ and\
  \bibinfo {author} {\bibfnamefont {M.~J.}\ \bibnamefont {White}},\ }\href
  {https://doi.org/10.1103/r5ps-pmh3} {\bibfield  {journal} {\bibinfo
  {journal} {Phys. Rev. D}\ }\textbf {\bibinfo {volume} {112}},\ \bibinfo
  {pages} {035022} (\bibinfo {year} {2025})},\ \Eprint
  {https://arxiv.org/abs/2506.01314} {arXiv:2506.01314 [hep-ph]} \BibitemShut
  {NoStop}%
\bibitem [{\citenamefont {Gargalionis}\ \emph
  {et~al.}(2026{\natexlab{b}})\citenamefont {Gargalionis}, \citenamefont
  {Moynihan}, \citenamefont {Trifinopoulos}, \citenamefont {Wallace},
  \citenamefont {White},\ and\ \citenamefont {White}}]{Gargalionis:2025iqs}%
  \BibitemOpen
  \bibfield  {author} {\bibinfo {author} {\bibfnamefont {J.}~\bibnamefont
  {Gargalionis}}, \bibinfo {author} {\bibfnamefont {N.}~\bibnamefont
  {Moynihan}}, \bibinfo {author} {\bibfnamefont {S.}~\bibnamefont
  {Trifinopoulos}}, \bibinfo {author} {\bibfnamefont {E.~N.~V.}\ \bibnamefont
  {Wallace}}, \bibinfo {author} {\bibfnamefont {C.~D.}\ \bibnamefont {White}},\
  and\ \bibinfo {author} {\bibfnamefont {M.~J.}\ \bibnamefont {White}},\ }\href
  {https://doi.org/10.1103/tcb9-3vpr} {\bibfield  {journal} {\bibinfo
  {journal} {Phys. Rev. D}\ }\textbf {\bibinfo {volume} {113}},\ \bibinfo
  {pages} {016007} (\bibinfo {year} {2026}{\natexlab{b}})},\ \Eprint
  {https://arxiv.org/abs/2508.14967} {arXiv:2508.14967 [hep-th]} \BibitemShut
  {NoStop}%
\bibitem [{\citenamefont {N{\'u}{\~n}ez}\ \emph {et~al.}(2026)\citenamefont
  {N{\'u}{\~n}ez}, \citenamefont {Pardina}, \citenamefont {Asorey},
  \citenamefont {Latorre},\ and\ \citenamefont
  {Cervera-Lierta}}]{Nunez:2025xds}%
  \BibitemOpen
  \bibfield  {author} {\bibinfo {author} {\bibfnamefont {C.}~\bibnamefont
  {N{\'u}{\~n}ez}}, \bibinfo {author} {\bibfnamefont {M.}~\bibnamefont
  {Pardina}}, \bibinfo {author} {\bibfnamefont {M.}~\bibnamefont {Asorey}},
  \bibinfo {author} {\bibfnamefont {J.~I.}\ \bibnamefont {Latorre}},\ and\
  \bibinfo {author} {\bibfnamefont {A.}~\bibnamefont {Cervera-Lierta}},\ }\href
  {https://doi.org/10.1103/4gyj-l8bj} {\bibfield  {journal} {\bibinfo
  {journal} {Phys. Rev. D}\ }\textbf {\bibinfo {volume} {113}},\ \bibinfo
  {pages} {096019} (\bibinfo {year} {2026})},\ \Eprint
  {https://arxiv.org/abs/2511.04358} {arXiv:2511.04358 [hep-th]} \BibitemShut
  {NoStop}%
\bibitem [{\citenamefont {Robin}\ and\ \citenamefont
  {Savage}(2025{\natexlab{a}})}]{Robin:2024oqc}%
  \BibitemOpen
  \bibfield  {author} {\bibinfo {author} {\bibfnamefont {C.~E.~P.}\
  \bibnamefont {Robin}}\ and\ \bibinfo {author} {\bibfnamefont {M.~J.}\
  \bibnamefont {Savage}},\ }\href {https://doi.org/10.1103/r8rq-y9tb}
  {\bibfield  {journal} {\bibinfo  {journal} {Phys. Rev. C}\ }\textbf {\bibinfo
  {volume} {112}},\ \bibinfo {pages} {044004} (\bibinfo {year}
  {2025}{\natexlab{a}})},\ \Eprint {https://arxiv.org/abs/2405.10268}
  {arXiv:2405.10268 [nucl-th]} \BibitemShut {NoStop}%
\bibitem [{\citenamefont {Robin}\ and\ \citenamefont
  {Savage}(2025{\natexlab{b}})}]{Robin:2025ymq}%
  \BibitemOpen
  \bibfield  {author} {\bibinfo {author} {\bibfnamefont {C.~E.~P.}\
  \bibnamefont {Robin}}\ and\ \bibinfo {author} {\bibfnamefont {M.~J.}\
  \bibnamefont {Savage}},\ }\href@noop {} {\bibinfo {title} {{Anti-Flatness and
  Non-Local Magic in Two-Particle Scattering Processes}}} (\bibinfo {year}
  {2025}{\natexlab{b}}),\ \Eprint {https://arxiv.org/abs/2510.23426}
  {arXiv:2510.23426 [quant-ph]} \BibitemShut {NoStop}%
\bibitem [{\citenamefont {Bai}(2023)}]{Bai:2023tey}%
  \BibitemOpen
  \bibfield  {author} {\bibinfo {author} {\bibfnamefont {D.}~\bibnamefont
  {Bai}},\ }\href {https://doi.org/10.1016/j.physletb.2023.138162} {\bibfield
  {journal} {\bibinfo  {journal} {Phys. Lett. B}\ }\textbf {\bibinfo {volume}
  {845}},\ \bibinfo {pages} {138162} (\bibinfo {year} {2023})},\ \Eprint
  {https://arxiv.org/abs/2306.04918} {arXiv:2306.04918 [nucl-th]} \BibitemShut
  {NoStop}%
\bibitem [{\citenamefont {Miller}(2023)}]{Miller:2023ujx}%
  \BibitemOpen
  \bibfield  {author} {\bibinfo {author} {\bibfnamefont {G.~A.}\ \bibnamefont
  {Miller}},\ }\href {https://doi.org/10.1103/PhysRevC.108.L031002} {\bibfield
  {journal} {\bibinfo  {journal} {Phys. Rev. C}\ }\textbf {\bibinfo {volume}
  {108}},\ \bibinfo {pages} {L031002} (\bibinfo {year} {2023})},\ \Eprint
  {https://arxiv.org/abs/2306.03239} {arXiv:2306.03239 [nucl-th]} \BibitemShut
  {NoStop}%
\bibitem [{\citenamefont {Cavallin}\ \emph {et~al.}(2026)\citenamefont
  {Cavallin}, \citenamefont {Thim},\ and\ \citenamefont
  {Forss{\'e}n}}]{Cavallin:2025kjn}%
  \BibitemOpen
  \bibfield  {author} {\bibinfo {author} {\bibfnamefont {A.~L.}\ \bibnamefont
  {Cavallin}}, \bibinfo {author} {\bibfnamefont {O.}~\bibnamefont {Thim}},\
  and\ \bibinfo {author} {\bibfnamefont {C.}~\bibnamefont {Forss{\'e}n}},\
  }\href {https://doi.org/10.1103/mlt1-z7t2} {\bibfield  {journal} {\bibinfo
  {journal} {Phys. Rev. C}\ }\textbf {\bibinfo {volume} {113}},\ \bibinfo
  {pages} {014005} (\bibinfo {year} {2026})},\ \Eprint
  {https://arxiv.org/abs/2510.09466} {arXiv:2510.09466 [nucl-th]} \BibitemShut
  {NoStop}%
\bibitem [{\citenamefont {Wita{\l}a}\ \emph {et~al.}(2025)\citenamefont
  {Wita{\l}a}, \citenamefont {Golak},\ and\ \citenamefont
  {Skibi{\'n}ski}}]{Witala:2025wvi}%
  \BibitemOpen
  \bibfield  {author} {\bibinfo {author} {\bibfnamefont {H.}~\bibnamefont
  {Wita{\l}a}}, \bibinfo {author} {\bibfnamefont {J.}~\bibnamefont {Golak}},\
  and\ \bibinfo {author} {\bibfnamefont {R.}~\bibnamefont {Skibi{\'n}ski}},\
  }\href {https://doi.org/10.1103/wr8m-4l77} {\bibfield  {journal} {\bibinfo
  {journal} {Phys. Rev. C}\ }\textbf {\bibinfo {volume} {112}},\ \bibinfo
  {pages} {044002} (\bibinfo {year} {2025})},\ \Eprint
  {https://arxiv.org/abs/2505.14401} {arXiv:2505.14401 [nucl-th]} \BibitemShut
  {NoStop}%
\bibitem [{\citenamefont {Wita{\l}a}(2026)}]{Witala:2025kat}%
  \BibitemOpen
  \bibfield  {author} {\bibinfo {author} {\bibfnamefont {H.}~\bibnamefont
  {Wita{\l}a}},\ }\href {https://doi.org/10.1140/epja/s10050-026-01791-x}
  {\bibfield  {journal} {\bibinfo  {journal} {Eur. Phys. J. A}\ }\textbf
  {\bibinfo {volume} {62}},\ \bibinfo {pages} {24} (\bibinfo {year} {2026})},\
  \Eprint {https://arxiv.org/abs/2510.10664} {arXiv:2510.10664 [nucl-th]}
  \BibitemShut {NoStop}%
\bibitem [{\citenamefont {Kirchner}\ \emph {et~al.}(2024)\citenamefont
  {Kirchner}, \citenamefont {Elkamhawy},\ and\ \citenamefont
  {Hammer}}]{Kirchner:2023dvg}%
  \BibitemOpen
  \bibfield  {author} {\bibinfo {author} {\bibfnamefont {T.}~\bibnamefont
  {Kirchner}}, \bibinfo {author} {\bibfnamefont {W.}~\bibnamefont
  {Elkamhawy}},\ and\ \bibinfo {author} {\bibfnamefont {H.-W.}\ \bibnamefont
  {Hammer}},\ }\href {https://doi.org/10.1007/s00601-024-01897-2} {\bibfield
  {journal} {\bibinfo  {journal} {Few Body Syst.}\ }\textbf {\bibinfo {volume}
  {65}},\ \bibinfo {pages} {29} (\bibinfo {year} {2024})},\ \Eprint
  {https://arxiv.org/abs/2312.14484} {arXiv:2312.14484 [nucl-th]} \BibitemShut
  {NoStop}%
\bibitem [{\citenamefont {Zanardi}\ \emph {et~al.}(2000)\citenamefont
  {Zanardi}, \citenamefont {Zalka},\ and\ \citenamefont
  {Faoro}}]{Zanardi:2000zz}%
  \BibitemOpen
  \bibfield  {author} {\bibinfo {author} {\bibfnamefont {P.}~\bibnamefont
  {Zanardi}}, \bibinfo {author} {\bibfnamefont {C.}~\bibnamefont {Zalka}},\
  and\ \bibinfo {author} {\bibfnamefont {L.}~\bibnamefont {Faoro}},\ }\href
  {https://doi.org/10.1103/PhysRevA.62.030301} {\bibfield  {journal} {\bibinfo
  {journal} {Phys. Rev. A}\ }\textbf {\bibinfo {volume} {62}},\ \bibinfo
  {pages} {030301} (\bibinfo {year} {2000})},\ \Eprint
  {https://arxiv.org/abs/quant-ph/0005031} {arXiv:quant-ph/0005031}
  \BibitemShut {NoStop}%
\bibitem [{\citenamefont {Nielsen}\ and\ \citenamefont
  {Chuang}(2012)}]{Nielsen:2012yss}%
  \BibitemOpen
  \bibfield  {author} {\bibinfo {author} {\bibfnamefont {M.~A.}\ \bibnamefont
  {Nielsen}}\ and\ \bibinfo {author} {\bibfnamefont {I.~L.}\ \bibnamefont
  {Chuang}},\ }\href {https://doi.org/10.1017/cbo9780511976667} {\emph
  {\bibinfo {title} {{Quantum Computation and Quantum Information}}}}\
  (\bibinfo  {publisher} {Cambridge University Press},\ \bibinfo {year}
  {2012})\BibitemShut {NoStop}%
\bibitem [{\citenamefont {Gottesman}(1998)}]{Gottesman:1998hu}%
  \BibitemOpen
  \bibfield  {author} {\bibinfo {author} {\bibfnamefont {D.}~\bibnamefont
  {Gottesman}},\ }in\ \href@noop {} {\emph {\bibinfo {booktitle} {{22nd
  International Colloquium on Group Theoretical Methods in Physics}}}}\
  (\bibinfo {year} {1998})\ pp.\ \bibinfo {pages} {32--43},\ \Eprint
  {https://arxiv.org/abs/quant-ph/9807006} {arXiv:quant-ph/9807006}
  \BibitemShut {NoStop}%
\bibitem [{\citenamefont {Aaronson}\ and\ \citenamefont
  {Gottesman}(2004)}]{Aaronson:2004xuh}%
  \BibitemOpen
  \bibfield  {author} {\bibinfo {author} {\bibfnamefont {S.}~\bibnamefont
  {Aaronson}}\ and\ \bibinfo {author} {\bibfnamefont {D.}~\bibnamefont
  {Gottesman}},\ }\href {https://doi.org/10.1103/PhysRevA.70.052328} {\bibfield
   {journal} {\bibinfo  {journal} {Phys. Rev. A}\ }\textbf {\bibinfo {volume}
  {70}},\ \bibinfo {pages} {052328} (\bibinfo {year} {2004})},\ \Eprint
  {https://arxiv.org/abs/quant-ph/0406196} {arXiv:quant-ph/0406196}
  \BibitemShut {NoStop}%
\bibitem [{\citenamefont {Leone}\ \emph {et~al.}(2022)\citenamefont {Leone},
  \citenamefont {Oliviero},\ and\ \citenamefont {Hamma}}]{Leone:2021rzd}%
  \BibitemOpen
  \bibfield  {author} {\bibinfo {author} {\bibfnamefont {L.}~\bibnamefont
  {Leone}}, \bibinfo {author} {\bibfnamefont {S.~F.~E.}\ \bibnamefont
  {Oliviero}},\ and\ \bibinfo {author} {\bibfnamefont {A.}~\bibnamefont
  {Hamma}},\ }\href {https://doi.org/10.1103/PhysRevLett.128.050402} {\bibfield
   {journal} {\bibinfo  {journal} {Phys. Rev. Lett.}\ }\textbf {\bibinfo
  {volume} {128}},\ \bibinfo {pages} {050402} (\bibinfo {year} {2022})},\
  \Eprint {https://arxiv.org/abs/2106.12587} {arXiv:2106.12587 [quant-ph]}
  \BibitemShut {NoStop}%
\bibitem [{\citenamefont {Wang}\ and\ \citenamefont {Li}(2023)}]{Wang:2023uog}%
  \BibitemOpen
  \bibfield  {author} {\bibinfo {author} {\bibfnamefont {Y.}~\bibnamefont
  {Wang}}\ and\ \bibinfo {author} {\bibfnamefont {Y.}~\bibnamefont {Li}},\
  }\href {https://doi.org/10.1007/s11128-023-04186-9} {\bibfield  {journal}
  {\bibinfo  {journal} {Quant. Inf. Proc.}\ }\textbf {\bibinfo {volume} {22}},\
  \bibinfo {pages} {444} (\bibinfo {year} {2023})}\BibitemShut {NoStop}%
\bibitem [{\citenamefont {Ohta}\ and\ \citenamefont
  {Sakurai}(2025)}]{Ohta:2025utz}%
  \BibitemOpen
  \bibfield  {author} {\bibinfo {author} {\bibfnamefont {M.}~\bibnamefont
  {Ohta}}\ and\ \bibinfo {author} {\bibfnamefont {K.}~\bibnamefont {Sakurai}},\
  }\href@noop {} {\bibinfo {title} {{Extremal Magic States from Symmetric
  Lattices}}} (\bibinfo {year} {2025}),\ \Eprint
  {https://arxiv.org/abs/2506.11725} {arXiv:2506.11725 [quant-ph]} \BibitemShut
  {NoStop}%
\bibitem [{\citenamefont {Liu}\ \emph {et~al.}(2026{\natexlab{b}})\citenamefont
  {Liu}, \citenamefont {Low},\ and\ \citenamefont {Yin}}]{Liu:2025frx}%
  \BibitemOpen
  \bibfield  {author} {\bibinfo {author} {\bibfnamefont {Q.}~\bibnamefont
  {Liu}}, \bibinfo {author} {\bibfnamefont {I.}~\bibnamefont {Low}},\ and\
  \bibinfo {author} {\bibfnamefont {Z.}~\bibnamefont {Yin}},\ }\href
  {https://doi.org/10.1088/2058-9565/ae3028} {\bibfield  {journal} {\bibinfo
  {journal} {Quantum Sci. Technol.}\ }\textbf {\bibinfo {volume} {11}},\
  \bibinfo {pages} {015035} (\bibinfo {year} {2026}{\natexlab{b}})},\ \Eprint
  {https://arxiv.org/abs/2502.17550} {arXiv:2502.17550 [quant-ph]} \BibitemShut
  {NoStop}%
\bibitem [{\citenamefont {Bethe}(1949)}]{Bethe:1949yr}%
  \BibitemOpen
  \bibfield  {author} {\bibinfo {author} {\bibfnamefont {H.~A.}\ \bibnamefont
  {Bethe}},\ }\href {https://doi.org/10.1103/PhysRev.76.38} {\bibfield
  {journal} {\bibinfo  {journal} {Phys. Rev.}\ }\textbf {\bibinfo {volume}
  {76}},\ \bibinfo {pages} {38} (\bibinfo {year} {1949})}\BibitemShut {NoStop}%
\bibitem [{\citenamefont {Schindler}\ \emph {et~al.}(2018)\citenamefont
  {Schindler}, \citenamefont {Singh},\ and\ \citenamefont
  {Springer}}]{Schindler:2018irz}%
  \BibitemOpen
  \bibfield  {author} {\bibinfo {author} {\bibfnamefont {M.~R.}\ \bibnamefont
  {Schindler}}, \bibinfo {author} {\bibfnamefont {H.}~\bibnamefont {Singh}},\
  and\ \bibinfo {author} {\bibfnamefont {R.~P.}\ \bibnamefont {Springer}},\
  }\href {https://doi.org/10.1103/PhysRevC.98.044001} {\bibfield  {journal}
  {\bibinfo  {journal} {Phys. Rev. C}\ }\textbf {\bibinfo {volume} {98}},\
  \bibinfo {pages} {044001} (\bibinfo {year} {2018})},\ \Eprint
  {https://arxiv.org/abs/1805.06056} {arXiv:1805.06056 [nucl-th]} \BibitemShut
  {NoStop}%
\bibitem [{\citenamefont {Ordonez}\ \emph {et~al.}(1996)\citenamefont
  {Ordonez}, \citenamefont {Ray},\ and\ \citenamefont {van
  Kolck}}]{Ordonez:1995rz}%
  \BibitemOpen
  \bibfield  {author} {\bibinfo {author} {\bibfnamefont {C.}~\bibnamefont
  {Ordonez}}, \bibinfo {author} {\bibfnamefont {L.}~\bibnamefont {Ray}},\ and\
  \bibinfo {author} {\bibfnamefont {U.}~\bibnamefont {van Kolck}},\ }\href
  {https://doi.org/10.1103/PhysRevC.53.2086} {\bibfield  {journal} {\bibinfo
  {journal} {Phys. Rev. C}\ }\textbf {\bibinfo {volume} {53}},\ \bibinfo
  {pages} {2086} (\bibinfo {year} {1996})},\ \Eprint
  {https://arxiv.org/abs/hep-ph/9511380} {arXiv:hep-ph/9511380} \BibitemShut
  {NoStop}%
\bibitem [{\citenamefont {Epelbaum}\ \emph {et~al.}(1998)\citenamefont
  {Epelbaum}, \citenamefont {Gloeckle},\ and\ \citenamefont
  {Meissner}}]{Epelbaum:1998ka}%
  \BibitemOpen
  \bibfield  {author} {\bibinfo {author} {\bibfnamefont {E.}~\bibnamefont
  {Epelbaum}}, \bibinfo {author} {\bibfnamefont {W.}~\bibnamefont {Gloeckle}},\
  and\ \bibinfo {author} {\bibfnamefont {U.-G.}\ \bibnamefont {Meissner}},\
  }\href {https://doi.org/10.1016/S0375-9474(98)00220-6} {\bibfield  {journal}
  {\bibinfo  {journal} {Nucl. Phys. A}\ }\textbf {\bibinfo {volume} {637}},\
  \bibinfo {pages} {107} (\bibinfo {year} {1998})},\ \Eprint
  {https://arxiv.org/abs/nucl-th/9801064} {arXiv:nucl-th/9801064} \BibitemShut
  {NoStop}%
\bibitem [{\citenamefont {Girlanda}\ \emph {et~al.}(2010)\citenamefont
  {Girlanda}, \citenamefont {Pastore}, \citenamefont {Schiavilla},\ and\
  \citenamefont {Viviani}}]{Girlanda:2010ya}%
  \BibitemOpen
  \bibfield  {author} {\bibinfo {author} {\bibfnamefont {L.}~\bibnamefont
  {Girlanda}}, \bibinfo {author} {\bibfnamefont {S.}~\bibnamefont {Pastore}},
  \bibinfo {author} {\bibfnamefont {R.}~\bibnamefont {Schiavilla}},\ and\
  \bibinfo {author} {\bibfnamefont {M.}~\bibnamefont {Viviani}},\ }\href
  {https://doi.org/10.1103/PhysRevC.81.034005} {\bibfield  {journal} {\bibinfo
  {journal} {Phys. Rev. C}\ }\textbf {\bibinfo {volume} {81}},\ \bibinfo
  {pages} {034005} (\bibinfo {year} {2010})},\ \Eprint
  {https://arxiv.org/abs/1001.3676} {arXiv:1001.3676 [nucl-th]} \BibitemShut
  {NoStop}%
\bibitem [{\citenamefont {Wigner}(1939)}]{Wigner:1939zz}%
  \BibitemOpen
  \bibfield  {author} {\bibinfo {author} {\bibfnamefont {E.~P.}\ \bibnamefont
  {Wigner}},\ }\href {https://doi.org/10.1103/PhysRev.56.519} {\bibfield
  {journal} {\bibinfo  {journal} {Phys. Rev.}\ }\textbf {\bibinfo {volume}
  {56}},\ \bibinfo {pages} {519} (\bibinfo {year} {1939})}\BibitemShut
  {NoStop}%
\bibitem [{\citenamefont {Hecht}\ and\ \citenamefont
  {Pang}(1969)}]{Hecht:1969ck}%
  \BibitemOpen
  \bibfield  {author} {\bibinfo {author} {\bibfnamefont {K.~T.}\ \bibnamefont
  {Hecht}}\ and\ \bibinfo {author} {\bibfnamefont {S.~C.}\ \bibnamefont
  {Pang}},\ }\href {https://doi.org/10.1063/1.1665007} {\bibfield  {journal}
  {\bibinfo  {journal} {J. Math. Phys.}\ }\textbf {\bibinfo {volume} {10}},\
  \bibinfo {pages} {1571} (\bibinfo {year} {1969})}\BibitemShut {NoStop}%
\bibitem [{\citenamefont {Li~Muli}\ \emph {et~al.}(2025)\citenamefont
  {Li~Muli}, \citenamefont {Dj{\"a}rv}, \citenamefont {Forss{\'e}n},\ and\
  \citenamefont {Phillips}}]{LiMuli:2025zro}%
  \BibitemOpen
  \bibfield  {author} {\bibinfo {author} {\bibfnamefont {S.~S.}\ \bibnamefont
  {Li~Muli}}, \bibinfo {author} {\bibfnamefont {T.~R.}\ \bibnamefont
  {Dj{\"a}rv}}, \bibinfo {author} {\bibfnamefont {C.}~\bibnamefont
  {Forss{\'e}n}},\ and\ \bibinfo {author} {\bibfnamefont {D.~R.}\ \bibnamefont
  {Phillips}},\ }\href@noop {} {\bibinfo {title} {{The role of spin-isospin
  symmetries in nuclear $\beta$-decays}}} (\bibinfo {year} {2025}),\ \Eprint
  {https://arxiv.org/abs/2503.16372} {arXiv:2503.16372 [nucl-th]} \BibitemShut
  {NoStop}%
\bibitem [{\citenamefont {Phillips}\ and\ \citenamefont
  {Schat}(2013)}]{Phillips:2013rsa}%
  \BibitemOpen
  \bibfield  {author} {\bibinfo {author} {\bibfnamefont {D.~R.}\ \bibnamefont
  {Phillips}}\ and\ \bibinfo {author} {\bibfnamefont {C.}~\bibnamefont
  {Schat}},\ }\href {https://doi.org/10.1103/PhysRevC.88.034002} {\bibfield
  {journal} {\bibinfo  {journal} {Phys. Rev. C}\ }\textbf {\bibinfo {volume}
  {88}},\ \bibinfo {pages} {034002} (\bibinfo {year} {2013})},\ \Eprint
  {https://arxiv.org/abs/1307.6274} {arXiv:1307.6274 [nucl-th]} \BibitemShut
  {NoStop}%
\bibitem [{\citenamefont {Pavon~Valderrama}\ and\ \citenamefont
  {Ruiz~Arriola}(2005)}]{PavonValderrama:2005ku}%
  \BibitemOpen
  \bibfield  {author} {\bibinfo {author} {\bibfnamefont {M.}~\bibnamefont
  {Pavon~Valderrama}}\ and\ \bibinfo {author} {\bibfnamefont {E.}~\bibnamefont
  {Ruiz~Arriola}},\ }\href {https://doi.org/10.1103/PhysRevC.72.044007}
  {\bibfield  {journal} {\bibinfo  {journal} {Phys. Rev. C}\ }\textbf {\bibinfo
  {volume} {72}},\ \bibinfo {pages} {044007} (\bibinfo {year}
  {2005})}\BibitemShut {NoStop}%
\bibitem [{\citenamefont {Stoks}\ \emph {et~al.}(1993)\citenamefont {Stoks},
  \citenamefont {Klomp}, \citenamefont {Rentmeester},\ and\ \citenamefont
  {de~Swart}}]{Stoks:1993tb}%
  \BibitemOpen
  \bibfield  {author} {\bibinfo {author} {\bibfnamefont {V.~G.~J.}\
  \bibnamefont {Stoks}}, \bibinfo {author} {\bibfnamefont {R.~A.~M.}\
  \bibnamefont {Klomp}}, \bibinfo {author} {\bibfnamefont {M.~C.~M.}\
  \bibnamefont {Rentmeester}},\ and\ \bibinfo {author} {\bibfnamefont {J.~J.}\
  \bibnamefont {de~Swart}},\ }\href {https://doi.org/10.1103/PhysRevC.48.792}
  {\bibfield  {journal} {\bibinfo  {journal} {Phys. Rev. C}\ }\textbf {\bibinfo
  {volume} {48}},\ \bibinfo {pages} {792} (\bibinfo {year} {1993})}\BibitemShut
  {NoStop}%
\bibitem [{\citenamefont {Weinberg}(1990)}]{Weinberg:1990rz}%
  \BibitemOpen
  \bibfield  {author} {\bibinfo {author} {\bibfnamefont {S.}~\bibnamefont
  {Weinberg}},\ }\href {https://doi.org/10.1016/0370-2693(90)90938-3}
  {\bibfield  {journal} {\bibinfo  {journal} {Phys. Lett. B}\ }\textbf
  {\bibinfo {volume} {251}},\ \bibinfo {pages} {288} (\bibinfo {year}
  {1990})}\BibitemShut {NoStop}%
\bibitem [{\citenamefont {Weinberg}(1991)}]{Weinberg:1991um}%
  \BibitemOpen
  \bibfield  {author} {\bibinfo {author} {\bibfnamefont {S.}~\bibnamefont
  {Weinberg}},\ }\href {https://doi.org/10.1016/0550-3213(91)90231-L}
  {\bibfield  {journal} {\bibinfo  {journal} {Nucl. Phys. B}\ }\textbf
  {\bibinfo {volume} {363}},\ \bibinfo {pages} {3} (\bibinfo {year}
  {1991})}\BibitemShut {NoStop}%
\end{thebibliography}%
\end{document}